\documentclass[ApJ]{aastex631}

\usepackage{color}
\graphicspath{{./}{figures/}}

\submitjournal{ApJ}
\received{June 15, 2021}
\accepted{August 12, 2021}
\shorttitle{A Previously Missed Planetary Mass Companion}

\begin{document}

\title{A Wide Planetary Mass Companion Discovered Through the Citizen Science Project Backyard Worlds: Planet 9 }

\correspondingauthor{J. Faherty}
\email{jfaherty@amnh.org}

\author[0000-0001-6251-0573]{Jacqueline K. Faherty}
\affil{Department of Astrophysics, American Museum of Natural History, Central Park West at 79th Street, NY 10024, USA }

\author[0000-0002-2592-9612]{Jonathan Gagn\'e}
\affiliation{Plan\'etarium Rio Tinto Alcan, Espace pour la vie, 4801 av. Pierre-De Coubertin, Montr\'eal, QC H1V~3V4, Canada}
\affiliation{Institute for Research on Exoplanets, Universit\'e de Montr\'eal, D\'epartement de Physique, C.P.~6128 Succ. Centre-ville, Montr\'eal, QC H3C~3J7, Canada} 

\author[0000-0001-9482-7794]{Mark Popinchalk}
\affil{Department of Astrophysics, American Museum of Natural History, Central Park West at 79th Street, NY 10024, USA }
\affiliation{Physics, The Graduate Center, City University of New York, New York, NY 10016, USA}
\affiliation{Department of Physics and Astronomy, Hunter College, City University of New York, 695 Park Avenue, New York, NY 10065, USA}

\author[0000-0003-0489-1528]{Johanna M. Vos}
\affil{Department of Astrophysics, American Museum of Natural History, Central Park West at 79th Street, NY 10024, USA }

\author[0000-0002-6523-9536]{Adam J. Burgasser}
\affil{Center for Astrophysics and Space Science, University of California San Diego, La Jolla, CA 92093, USA}

\author[0000-0002-7587-7195]{J{\"o}rg Sch{\"u}mann}
\affil{Backyard Worlds: Planet 9, USA}

\author[0000-0002-6294-5937]{Adam C. Schneider}
\affil{United States Naval Observatory, Flagstaff Station, 10391 West Naval Observatory Rd., Flagstaff, AZ 86005, USA}
\affil{Department of Physics and Astronomy, George Mason University, MS3F3, 4400 University Drive, Fairfax, VA 22030, USA}

\author[0000-0003-4269-260X]{J. Davy Kirkpatrick}
\affil{IPAC, Mail Code 100-22, Caltech, 1200 E. California Blvd., Pasadena, CA 91125, USA}

\author[0000-0002-1125-7384]{Aaron M. Meisner}
\affil{NSF's National Optical-Infrared Astronomy Research Laboratory, 950 N. Cherry Ave., Tucson, AZ 85719, USA}

\author[0000-0002-2387-5489]{Marc J. Kuchner}
\affiliation{NASA Goddard Space Flight Center, Exoplanets and Stellar Astrophysics Laboratory, Code 667, Greenbelt, MD 20771}

\author[0000-0001-8170-7072]{Daniella C. Bardalez Gagliuffi}
\affiliation{Department of Astrophysics, American Museum of Natural History, Central Park West at 79th Street, NY 10024, USA }

\author[0000-0001-7519-1700]{Federico Marocco}
\affil{IPAC, Mail Code 100-22, Caltech, 1200 E. California Blvd., Pasadena, CA 91125, USA}

\author[0000-0001-7896-5791]{Dan Caselden}
\affil{Gigamon Applied Threat Research, 619 Western Avenue, Suite 200, Seattle, WA 98104, USA}

\author[0000-0003-4636-6676]{Eileen C. Gonzales}
\affil{Department of Astronomy and Carl Sagan Institute, Cornell University, 122 Sciences Drive, Ithaca, NY 14853, USA}
\affil{51 Pegasi b Fellow}

\author[0000-0003-4083-9962]{Austin Rothermich}
\affil{Department of Astrophysics, American Museum of Natural History, Central Park West at 79th Street, NY 10024, USA }

\author[0000-0003-2478-0120]{Sarah L. Casewell}
\affiliation{Department of Physics and Astronomy, University of Leicester, University Road, Leicester LE1 7RH, UK}

\author[0000-0002-1783-8817]{John H. Debes}
\affil{ESA for AURA, Space Telescope Science Institute, 3700 San Martin Drive, Baltimore, MD 21218, USA
}

\author[0000-0003-2094-9128]{Christian Aganze}
\affil{Center for Astrophysics and Space Science, University of California San Diego, La Jolla, CA 92093, USA}

\author{Andrew Ayala}
\affil{Department of Astrophysics, American Museum of Natural History, Central Park West at 79th Street, NY 10024, USA }

\author[0000-0002-5370-7494]{Chih-Chun Hsu}
\affil{Center for Astrophysics and Space Science, University of California San Diego, La Jolla, CA 92093, USA}

\author[0000-0003-3501-8967]{William J. Cooper}
\affil{Centre for Astrophysics Research, University of Hertfordshire, Hatfield, AL10 9AB, United Kingdom}
\affil{Istituto Nazionale di Astrofisica, Osservatorio Astrofisico di Torino, Strada Osservatorio 20, 10025 Pino Torinese, Italy}

\author[0000-0002-4424-4766]{R. L. Smart}
\affil{Istituto Nazionale di Astrofisica, Osservatorio Astrofisico di Torino, Strada Osservatorio 20, 10025 Pino Torinese, Italy}

\author[0000-0003-0398-639X]{Roman Gerasimov}
\affil{Center for Astrophysics and Space Science, University of California San Diego, La Jolla, CA 92093, USA}

\author[0000-0002-9807-5435]{Christopher A. Theissen}
\altaffiliation{NASA Sagan Fellow}
\affil{Center for Astrophysics and Space Science, University of California San Diego, La Jolla, CA 92093, USA}

\author{The Backyard Worlds: Planet 9 Collaboration}

 \begin{abstract}
Through the Backyard Worlds: Planet 9 citizen science project we discovered a late-type L dwarf co-moving with the young K0 star BD+60 1417 at a projected separation of 37$\arcsec$ or 1662 AU.  The secondary -- CWISER J124332.12+600126.2 (W1243) -- is detected in both the CatWISE2020 and 2MASS reject tables.  The photometric distance and CatWISE proper motion both match that of the primary within $\sim$1$\sigma$ and our estimates for chance alignment yield a zero probability. Follow-up near infrared spectroscopy reveals W1243 to be a very red 2MASS ($J$-$K_{s}$=2.72), low-surface gravity source that we classify as L6 - L8$\gamma$.  Its spectral morphology strongly resembles that of confirmed late-type L dwarfs in 10 - 150 Myr moving groups as well as that of planetary mass companions.  The position on near- and mid-infrared color-magnitude diagrams indicates the source is redder and fainter than the field sequence, a telltale sign of an object with thick clouds and a complex atmosphere.  For the primary we obtained new optical spectroscopy and analyzed all available literature information for youth indicators.  We conclude that the Li I abundance, its loci on color-magnitude and color-color diagrams, and the rotation rate revealed in multiple TESS sectors are all consistent with an age of 50 - 150 Myr.  Using our re-evaluated age of the primary, the Gaia parallax along with the photometry and spectrum for W1243 we find a $T_{\rm eff}$=1303$\pm$31 K, logg=4.3$\pm$0.17 cm s$^{-2}$, and a mass of 15$\pm$5 M$_{Jup}$.  We find a physical separation of $\sim$1662 AU and a mass ratio of $\sim$0.01 for this system. Placing it in context with the diverse collection of binary stars, brown dwarf and planetary companions, the BD+60 1417 system falls in a sparsely sampled area where the formation pathway is difficult to assess.

\end{abstract}


\section{Introduction\label{sec:intro}}
Young stars near the Sun are the targeting ground for direct imaging campaigns in search of hot massive planets.  Over the past two decades projects such as the Gemini NICI planet finding campaign, The Gemini Planet Imager Exoplanet Survey, the VLT/SPHERE survey for exoplanets, the Lyot Project Direct Imaging Survey of Substellar Companions, among others have examined the space around hundreds of young hot stars (\citealt{Chauvin2015}, \citealt{Vigan2020}, \citealt{Biller2013}, \citealt{Macintosh2008}, \citealt{Leconte2010}).  Interestingly the number of giant exoplanets --and even brown dwarfs -- at 10-100 AU separations from their host stars has been far fewer than hoped or predicted (\citealt{Nielsen2019}).  

Even still there are now 37 systems listed on the NASA exoplanet archive\footnote{https://exoplanetarchive.ipac.caltech.edu/ As of April 23 2021 there are 4383 confirmed planets}  site as direct imaging discoveries of a planetary mass companion (using the 13M$_{Jup}$ definition) where the total mass of the system is larger than 0.1 $M_{\sun}$.  Among those 37 systems are the first multiple imaged exoplanet system  (HR8799bcde; \citealt{Marois2008,Marois2010}), systems with brown dwarf or extremely low mass star hosts (e.g. 2M1207b; \citealt{Chauvin04} and 2M0219b; \citealt{Artigau15}), and companions still accreting material (e.g. PDS70bc; \citealt{Muller2018}).  There is a large grey area in our understanding of how these systems formed given their largely model dependent masses and wide range of separations.  Competing pathways for the formation of these systems include the more traditional mechanisms of core accretion and disk instability for planets and cloud fragmentation for brown dwarfs.  The ability to differentiate between formation mechanism would aid in creating well defined classes of brown dwarfs versus exoplanets. While this is a well-trodden exercise (e.g. \citealt{Schlaufman2018}, \citealt{Brandt2014}, \citealt{Metchev08}, \citealt{Raghavan2010}, \citealt{Bowler2020b}) only tentative conclusions have been drawn on how to differentiate between a disk-formed object and a cloud fragmented object.

Meanwhile, numerous isolated low surface gravity brown dwarfs have been detected and confirmed to be members of 10 - 200 Myr moving groups like Tucana Horologium, $\beta$ Pictoris, and AB Doradus (e.g. \citealt{Gagne18, Gagne15}, \citealt{Best2017}, \citealt{Faherty13, Faherty16}, \citealt{Liu13}, \citealt{Zhang2021}, \citealt{Schneider2018}).  These sources are distinctly different from field brown dwarfs in their spectral morphology, colors, and absolute magnitudes (\citealt{Faherty16}, \citealt{Liu16}).  Studies of their photometric and spectral variability allude to complex atmospheres with strong weather related phenomena (e.g \citealt{Metchev2015}, \citealt{Vos2020}).  Given their young ages and low temperatures, many have masses that overlap with those of directly imaged planetary mass companions at just above and below 13 M$_{Jup}$ (see for example \citealt{Gagne17TWHya}).

In many ways the populations of giant exoplanets and isolated young brown dwarfs are identical. There are, at present, no defined spectral characteristics imprinted on a source that indicate which formation mechanism was responsible for its creation. Studies have shown that planetary mass companions and isolated planetary mass sources share spectral morphology as well as positions on color-magnitude diagrams (e.g. \citealt{Faherty16}, \citealt{Gagne15}, \citealt{Liu16}).  They have identical weather related phenomena (e.g. \citealt{Vos2019}, \citealt{Biller2018}, \citealt{Zhou2016}), temperatures, and gravities; therefore, it benefits both populations if they are studied in tandem.  

In this paper we report the discovery of a new system containing a planetary mass companion.  Using the Backyard Worlds: Planet 9 citizen science project, we found that the young K0 star BD+60 1417 was co-moving with a late-type L dwarf with hallmark signatures of youth.  In section~\ref{sec:Discovery} we discuss the discovery of the system, in section~\ref{sec:newdata} we detail new data taken on both the primary and the secondary, and in section~\ref{sec:components} we provide a detailed discussion on each component. Section~\ref{sec:W1243spectra} discusses the spectrum of W1243 and section~\ref{sec:chancealign} evaluates the probability of chance alignment for the system. Section~\ref{sec:CMD} evaluates W1243 on color-magnitude diagrams, and section~\ref{sec:age} re-evaluates the age of the primary BD+60 1417. Section~\ref{sec:fundamentals} details how fundamental parameters were calculated for each component and section~\ref{sec:characteristics} places the BD+60 1417 system in context with other planetary mass companion systems.  Section~\ref{sec:discussion} is a discussion on potential formation mechanisms for this system.  Conclusions are presented in section~\ref{sec:conclusions}.

\begin{figure*}
\begin{center}
\includegraphics[width=0.8\linewidth]{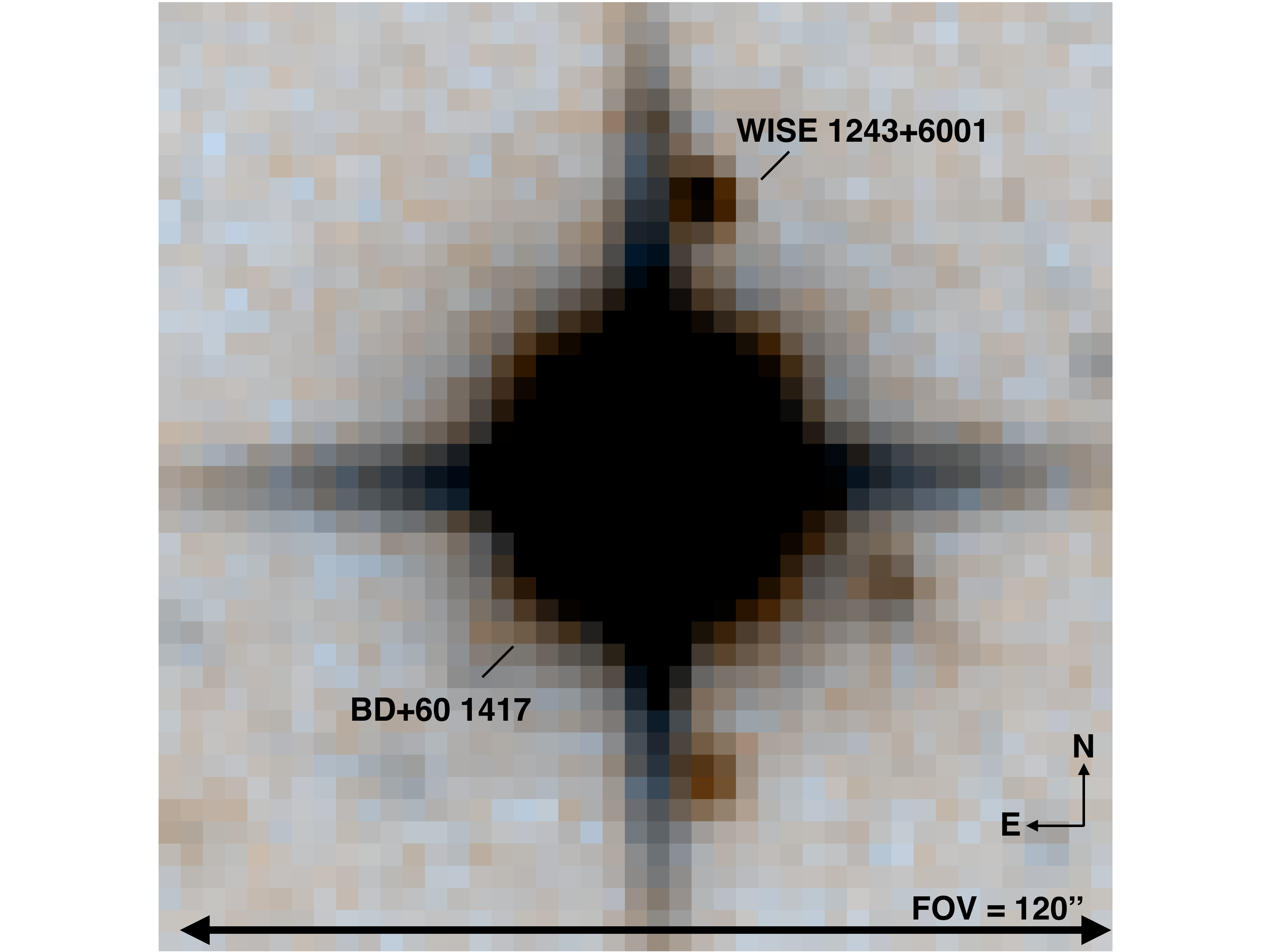}
\caption{The finder chart for the BD+60 1417 and W1243+6001 system taken from the WiseView website (\citealt{Caselden18}).   To see the animated motion between available $WISE$ epochs visit the URL http://byw.tools/wiseview-v2 and use coordinates RA,DEC=190.88751026191, 60.01435471736.}
\label{fig:finder}
\end{center}
\end{figure*}

\section{Discovery\label{sec:Discovery}}
The Backyard Worlds: Planet 9 citizen science project (Backyard Worlds for short) has been operational since February 2017. The scientific goal of the project is to complete the census of the solar neighborhood (including the solar system, e.g. Planet 9) with objects that are detectable primarily at mid infrared wavelengths and that were missed by previous searches (see \citealt{Kuchner17}, \citealt{Debes19}, \citealt{Faherty2020}, \citealt{Bardalez2020}, \citealt{Meisner2020}, \citealt{Schneider2020}, \citealt{Kirkpatrick2021}, \citealt{Rothermich2021}, \citealt{Jalowiczor2021}). Backyard Worlds utilizes multiple epochs of NASA's Wide-field Infrared Survey Explorer ($WISE$) mission at both W1 ($\sim$3.4 $\mu$m) and W2 ($\sim$4.6$\mu$m) wavelengths. Project participants are asked to blink between four unWISE images (see \citealt{Meisner17b}) where the time-span between the first and last image is $\sim$ 4.5 years. Given this time baseline, objects of significant motion (e.g. $>$ 200\,mas\,yr$^{-1}$) are relatively easy to visually identify.

The BackyardWorlds.org\footnote{http://www.backyardworlds.org} website hosted by Zooniverse provides multiple avenues for reporting a proper motion candidate of scientific interest to the research team (see \citealt{Kuchner17} for details).  Since the program began a significant number of Backyard Worlds users have become heavily involved with the research aspect of the project and have earned the title ``super users".  These participants attend weekly calls with the science team, have increased contact via email and are much more engaged than a casual online subscriber to the project.  With the increased mentorship from the research team, the super users tend to take on more complex searches outside of the website itself and as such have well tuned eyes to detecting more obscure targets.  In fact one citizen scientist, Dan Caselden--also co-author on this work--, has developed a tool called WiseView (\citealt{Caselden18}) which allows users to flip through available $WISE$ images without the image subtraction used on the Zooniverse site.  WiseView has an array of functionality (changing the image stretch, zooming in and out, overlaying Gaia astrometry, etc.) which compliments the Zooniverse site but also greatly aids in complex searches.

For this discovery, citizen scientist J{\"o}rg Sch{\"u}mann was searching through subsets of images in WiseView and identified a faint comoving system. In this case, the secondary was both in the halo of the primary and along one of its diffraction spikes therefore it was easily missed by sophisticated catalog searches. Visual inspection was critical to the discovery of the secondary and it is an exemplary case of the power of the Backyard Worlds project. 

J{\"o}rg used the Google form as well as the Backyard Worlds email distribution list to alert researchers to the discovery.  On 10 April 2018, the motion of W1243 was vetted by the research team and added to the high priority follow-up target list with a note that it was potentially co-moving with BD+60 1417. Figure~\ref{fig:finder} shows a screenshot from the WiseView website (\citealt{Caselden18}) which was used to identify and confirm the system. 

\section{New Data\label{sec:newdata}}
To characterize this potential co-moving system, we followed up on both the primary and secondary with optical and infrared spectrographs (respectively).  

\subsection{Optical Spectroscopy}
On the night of 10 January 2021 (UT) we observed the primary BD+60 1417 using the Kast optical spectrograph on the Shane 3m telescope at Lick observatory. Conditions were variable with some wind and cirrus clouds. The average seeing at the time of observations was $\sim$2$\arcsec$. We observed the system using the 2$\arcsec$ slit with one 60s exposure in the red channel (6300 - 9000 $\AA$).  The A0V telluric star 81 UMa was observed for absorption calibration and the flux standard Hiltner~600 was observed during the night for flux calibration \citep{hamuy1994}. We also obtained flat field and He, Hg-A, Ne, and Spare Ar lamps at the start of the night for pixel response and wavelength calibration. Data were reduced using the \texttt{kastredux} package\footnote{\url{https://github.com/aburgasser/kastredux}} using default settings.

\begin{figure*}
\begin{center}
\includegraphics[width=0.8\linewidth]{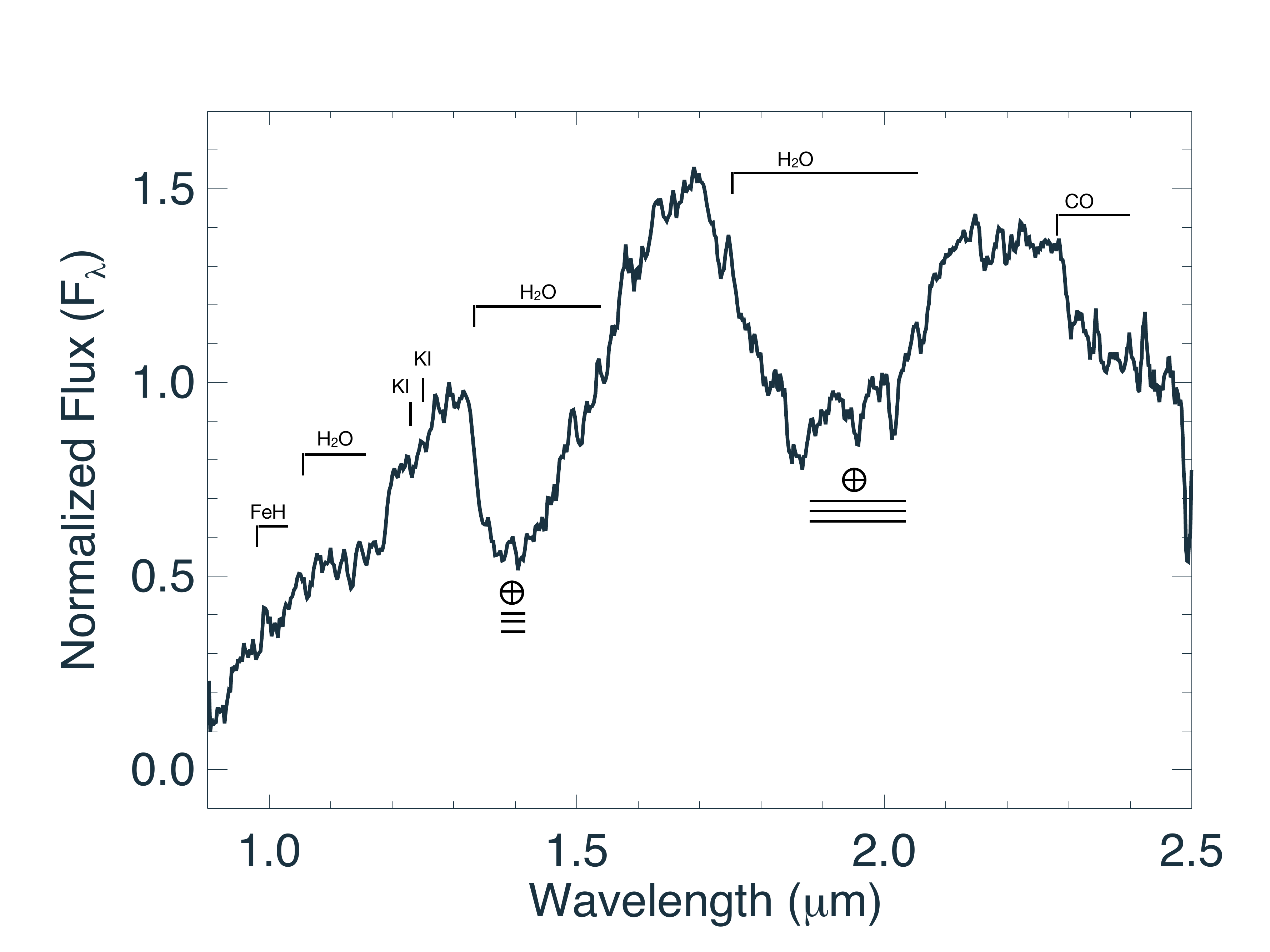}
\caption{The SpeX Prism spectrum for W1243 normalized over the $J$ band with prominent features labeled.}
\label{fig:spectrasecondary}
\end{center}
\end{figure*}

\subsection{SpeX Prism Spectroscopy}
We observed the secondary W1243 on two separate nights using the SpeX spectrograph on NASA's IRTF telescope.  The first data were taken on the night of 04 January 2021 (UT) under good conditions with minimal cloud coverage and the second on 31 January 2021 (UT) under varying cloud conditions.  The data were taken in prism mode using the 0.8$\arcsec$ slit to achieve a resolving power of $\sim$100 - 500 over the 0.8 - 2.5$\mu$m coverage. We observed the object on two separate nights to get ample signal as the first spectrum was taken close to dusk and therefore we could not integrate for our desired amount of time.  On 04 January we obtained 8 AB nods using 180s exposures on the target and then acquired the A0 star HD 99966 for telluric correction using 0.1s exposures and 20 AB nods.  On 31 January we obtained 28 AB nods using 180s exposures on the target and then acquired the A0 star HD 116405 for telluric correction using 0.1s exposures and 20 AB nods.  All data were reduced using the \texttt{Spextool} package (\citealt{Cushing04}) with telluric correction and flux calibration of the A0 stars following the technique described in \citet{Vacca03}.   

\begin{figure*}
\begin{center}
\includegraphics[width=0.8\linewidth]{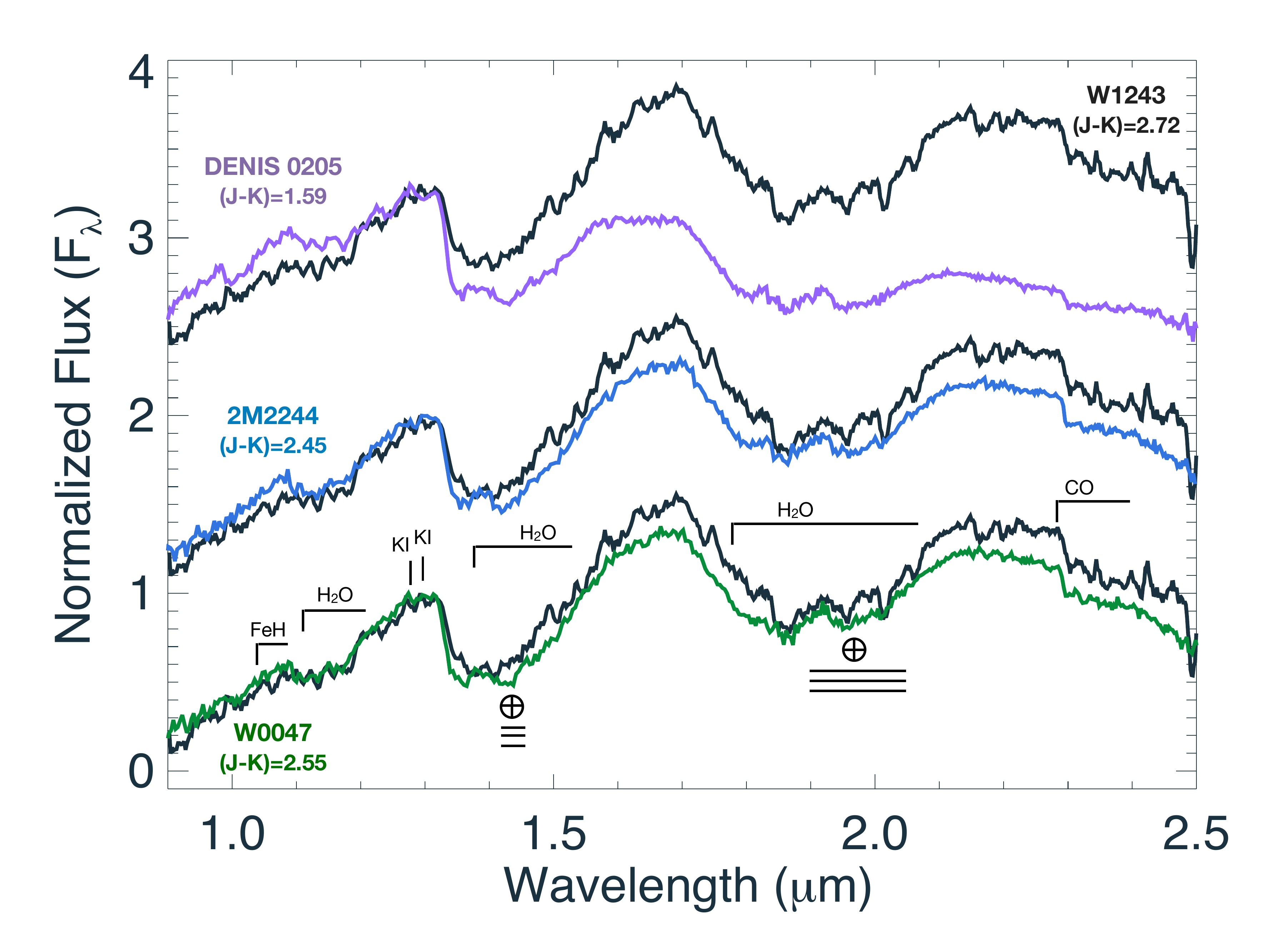}
\caption{The SpeX Prism spectra for W1243 as compared to data on the field age optical L7 standard DENIS 0205, and the $\sim$ 150 Myr AB Doradus L dwarfs 2M2244 and W0047.  Under the name of each object we list the 2MASS ($J$-$K_{s}$) color. Relevant spectral features are labeled.}
\label{fig:ComparisonPanel}
\end{center}
\end{figure*}

\section{Details on the components\label{sec:components}}
\subsection{Primary}
BD+60 1417 is a well known nearby star (e.g. \citealt{Tycho}).  Given the notable X-ray detection in \citet{Rosat} and the Li I absorption for this source it was an attractive target for numerous nearby young star studies (e.g., \citealt{Wichmann2003}).  \citet{Wichmann2003} estimate an age for the star of 50 - 150 Myr based on its X-ray flux compared to equivalent Pleiades sources. Due to its relatively young age, nearby distance ($\sim$45\,pc) and K0 spectral type (solar type-esque), BD+60 1417 was a target of choice for the Gemini deep planet survey (\citealt{Lafreniere2007}), the Palomar/Keck adaptive optics survey of young solar analogs (\citealt{Metchev2009}) as well as other planet finding programs (e.g. \citealt{Heinze2010}).  There was an object of interest imaged around BD+60 1417 by \citet{Lafreniere2007} and \citet{Heinze2010} but it was ruled out as a background object due to a mismatched proper motion.  As such BD+60 1417 remained a young K0 star with no directly imaged exoplanet discovered to date and became a statistical null-detection point for analysis papers looking at the frequency of giant planets (e.g. \citealt{Nielsen2010}, \citealt{Vigan2017}, \citealt{Meshkat2017}).  

\begin{figure}[!ht]
\begin{center}
$\begin{array}{cc}

\includegraphics[width=3.1in]{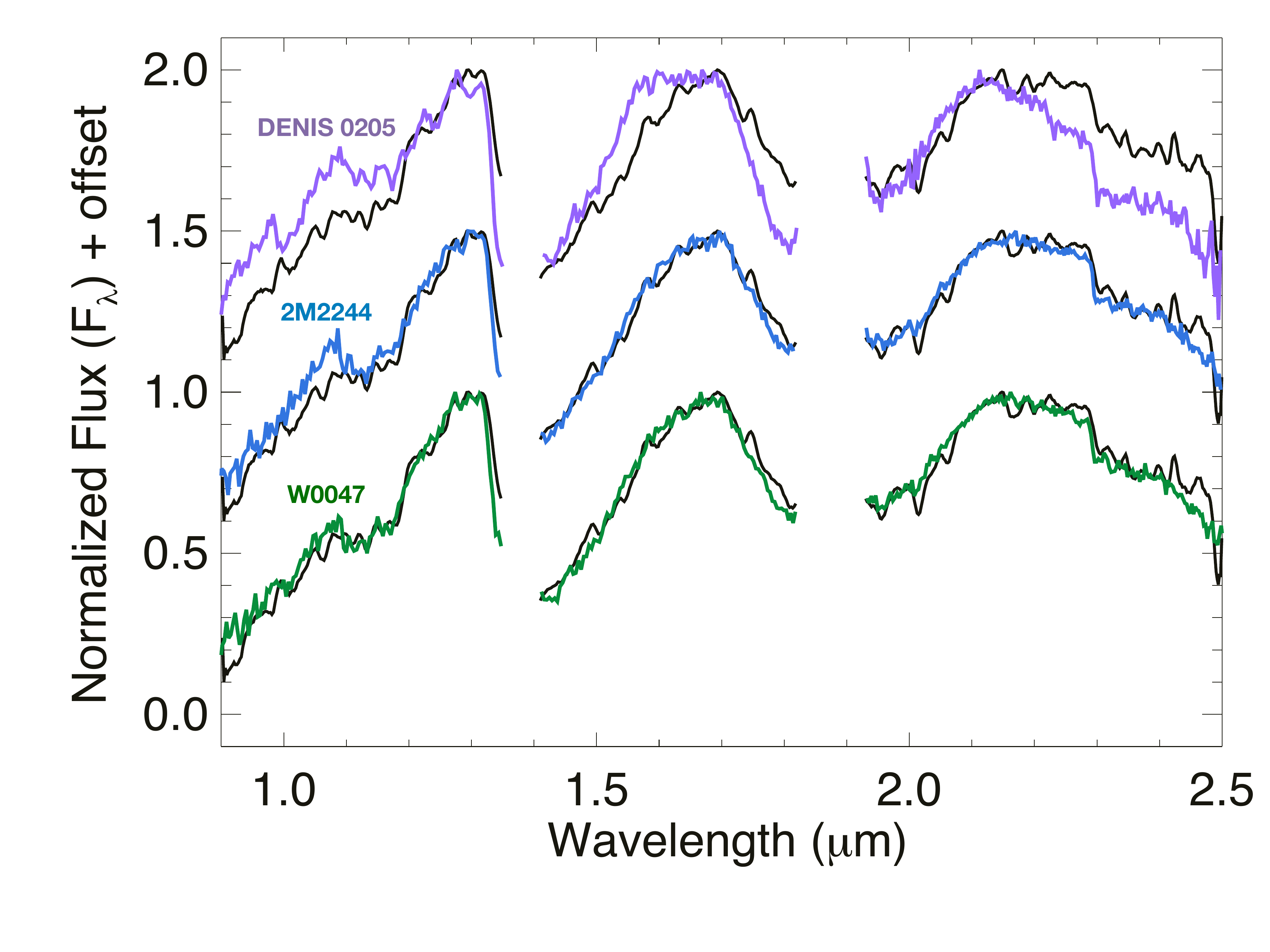}&\\
\includegraphics[width=3.1in]{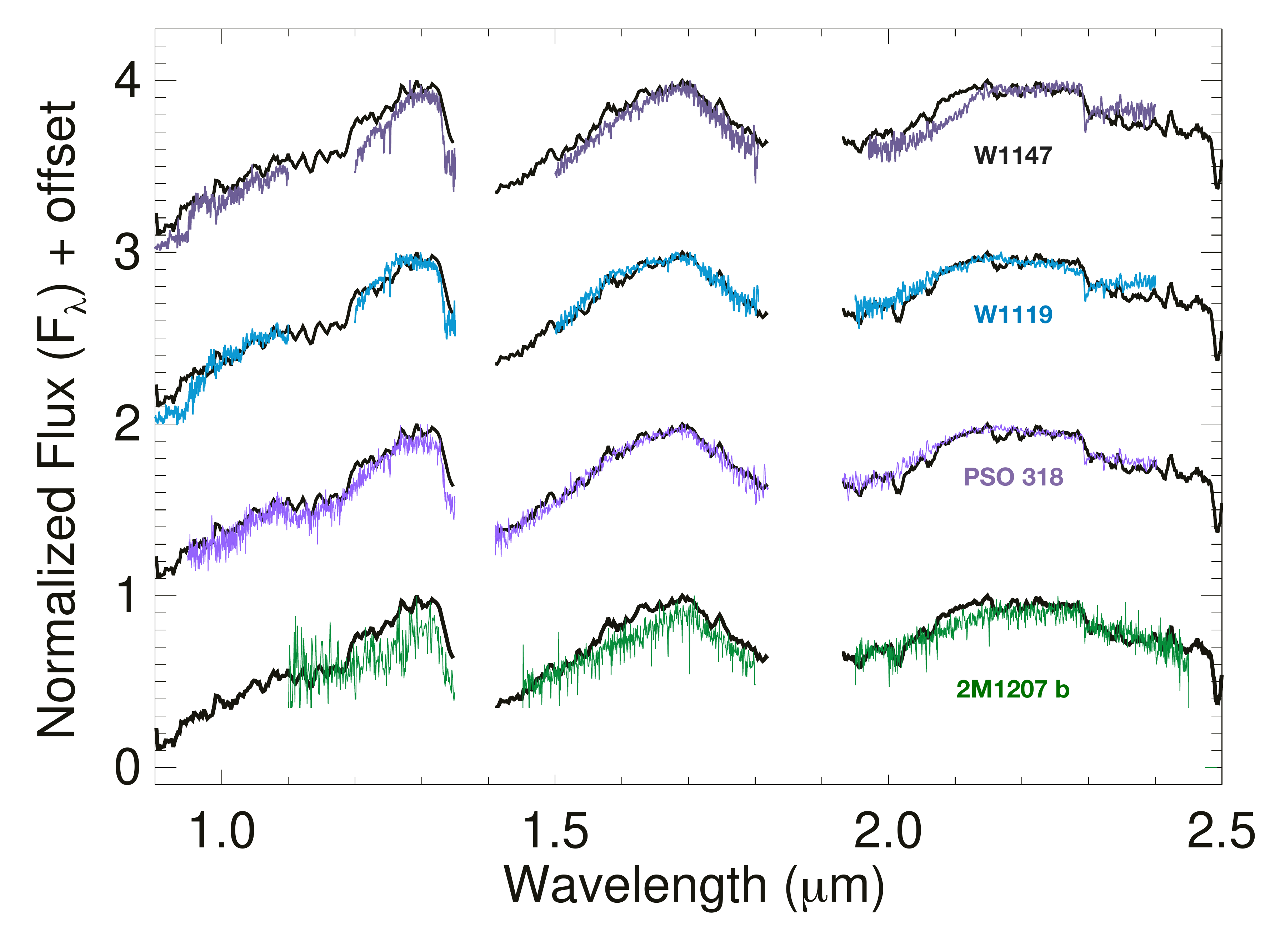} \\
\includegraphics[width=3.1in]{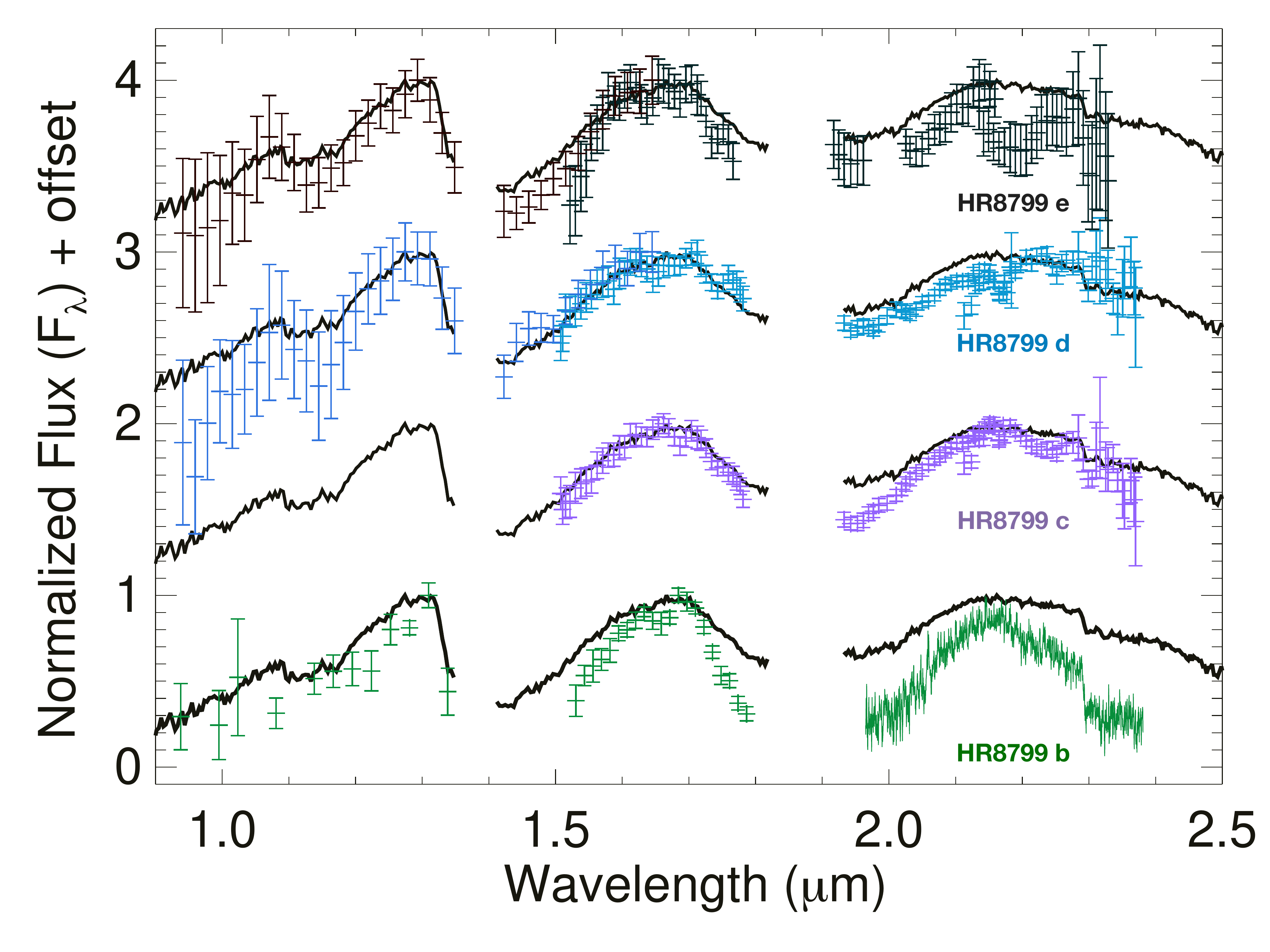} \\
\end{array}$
\end{center}
\caption{ The SpeX prism spectrum of W1243 normalized band by band and compared to various isolated young L dwarfs and planetary mass companions.  Top: Compared to DENIS 0205, 2M2244, and W0047.  Middle: Compared to 2M1207b, PSO318, W1147, and W1119. Bottom: Compared to the HR8799bcde planets. Error bars are shown for the HR8799 companions given their significance against the data. }
\label{fig:spectra-bands}
\end{figure}

\subsection{Secondary\label{sec:secondary}}
CWISER  J124332.12+600126.2 (W1243) is detected in both the 2MASS Survey Point Source Reject (using a reliability limit $>$ 20\%) and CatWISE2020 reject tables (\citealt{Cutri03}, \citealt{Catwise2020}). For 2MASS there was a reported $K_{s}$ photometric point while $J$ and $H$ bands were limits only.  W1243 falls in a part of the sky observed by Pan-STARRS although only a detection in $y$ band is reported. Despite being in the CatWISE2020 reject table, W1243 is well detected at both W1 and W2 bands.  Looking at the flags, it is noted as being near a diffraction spike and near the halo of a bright star in  both W1 and W2 bands.  Proper motion component values for W1243 are listed in the CatWISE2020 reject table.  While the primary has been looked over by numerous surveys for a close-in companion with a comparable or lower temperature to W1243, no survey looked far enough from the primary to detect this object.

\section{Analysis of the NIR spectrum of W1243\label{sec:W1243spectra}}
The SpeX prism spectrum of W1243 is shown in Figure~\ref{fig:spectrasecondary} normalized over the peak of the $J$ band with pertinent atomic and molecular features labeled. The source shows strong H$_{2}$O and CO absorption as well as FeH, and KI features.  The distribution of flux across the normalized spectrum visually indicates a very red late-type L dwarf.  We used the 2MASS filter profiles overlaid on the SpeX spectrum to compute synthetic magnitudes for this source in $J$ and $H$.  We calibrated using the synthetic ($J$- $H$) and ($J$-$K_{s}$) colors along with the 2MASS $K_{s}$ detection and list the photometry in Table~\ref{tab:param}. We find the ($J$-$K_{s}$) color of 2.72 makes W1243 one of the reddest known brown dwarfs characterized to date (see e.g. \citealt{Marocco2014} and \citealt{Liu13} for details on ULAS J222711-004547 and PSO 318 the two record holders for reddest isolated brown dwarfs).

Figure ~\ref{fig:ComparisonPanel} shows W1243 compared to three late-type L dwarfs normalized over the peak of the $J$ band.  We compare W1243 to the $\sim$150 Myr AB Doradus moving group objects WISEP J004701.06+680352.1 (W0047, \citealt{Gizis2015}) and 2MASS J22443167+2043433 (2M2244, \citealt{Faherty16}, \citealt{Vos2018}) as well as the field source and optical L7 standard DENIS-P J0205.4-1159 (DENIS 0205, \citealt{Kirkpatrick99}).  Under each object name we also denote the 2MASS ($J$-$K$) color.  The spectrum of W1243 is substantially redder than DENIS 0205.  It most closely resembles W0047 in its spectral morphology although it also appears slightly redder than this source. 

W0047 was evaluated by \citet{Gizis2015} to be an infrared L7 with signatures of a low surface gravity. \citet{Faherty16} evaluated a sample of over 150 low surface gravity M and L dwarfs and determined that both W0047 and 2M2244 should be considered L6 - L8$\gamma$ prototypes where the $\gamma$ designation indicates a very low surface gravity (see \citealt{Kirkpatrick2005}, \citealt{Cruz09}, \citealt{Gagne15}). 

Evaluating the exact spectral subtype of a red L dwarf against the array of known objects can be difficult given the significant color difference across the spectrum.  As recommended by \citet{Cruz18} we also performed a band by band normalization across $J$, $H$, and $K_{s}$ so we could compare to other field, young and planetary mass objects -- the latter of which only have spectra taken band by band.  

The top panel of Figure~\ref{fig:spectra-bands} shows the band by band normalization of W1243 as compared to DENIS 0205, 2M2244, and W0047.  With the color term minimized, it is far easier to compare the spectral morphology.  Most notably, compared to the field source DENIS 0205, W1243 has a much sharper $H$ band peak which is a hallmark signature of low surface gravity L dwarfs (e.g. \citealt{Allers13}).  The $K$ band of W1243 is also enhanced compared to DENIS 0205, a feature seen in other low gravity L dwarfs and attributed to a change in collision induced $H_{2}$ absorption (e.g. \citealt{Saumon2012}).  The depth of the NaI and KI alkali lines in the $J$ band is also a low surface gravity diagnostic.  \citet{Allers13} and \citet{Martin17} evaluated the equivalent width values from medium resolution SpeX or NIRSPEC (respectively) data to determine trends between gravity and alkali depths.  In general, younger sources show narrow, shallow lines  compared to older field objects.  The low resolution SpeX prism data hint that the KI lines are shallower for W1243 than DENIS 0205 and far more similar to W0047 or 2M2244.  

We also investigated the recommended indices from \citet{Allers13} to evaluate the gravity class for W1243.  That work was calibrated on sources warmer than W1243 (L5 or earlier) therefore only the $H$- band continuum index can be reliably used.  Using that we find this source would be classified as a  L6$\gamma$ consistent with our analysis.   

Based on these spectral comparisons we determine that W1243 should be considered a very low surface gravity late-type L dwarf and we assign a spectral type of L6 - L8$\gamma$ mimicking the conclusions for both W0047 and 2M2244 which are members of the $\sim$150 Myr AB Doradus moving group.  

We expanded our analysis to include other L6 - L8$\gamma$ objects which are in younger moving groups.  In the middle panel of Figure~\ref{fig:spectra-bands} we do a band by band comparison of W1243 to the candidate $\sim$10 Myr TW Hya association L dwarfs WISEA J114724.10-204021.3 (W1147; \citealt{Schneider2016}) and WISEA J111932.43-113747.7 (\citealt{Kellogg2016}) as well as the $\sim$24 Myr $\beta$ Pictoris moving group member PSO J318.5338-22.8603 (PSO 318; \citealt{Liu13}).  The spectra for W1147 and W1119 were obtained from FIRE Echelle (from \citealt{Faherty16}) and are  higher resolution than the prism data (R$\sim$3000-5000 compared to R$\sim$100-500 for prism). Moreover, they are stitched together order by order which might introduce shape anomalies across a band (e.g. W1147 K band). In this comparison we find that W1243 shows similarities to all three of the young sources with W1147 showing more of a triangular H band and W1119 and PSO 318 showing sharper $J$ band peaks.  However it is well known that even low surface gravity L dwarfs in the same association with the same estimated temperatures can show a range in their spectral features (e.g., 2M0355, CD35 -2722B from AB Doradus moving group \citealt{Allers13}).  Consequently we can not conclude whether W1243 is better matched by an AB Doradus moving group age of 149$^{+51}_{-19}$ {Myr} or a TW Hya or $\beta$ Pictoris younger age of 10$\pm$3, $<$24$\pm$3 Myr respectively (\citealt{Bell2015}).  Based on these comparisons we can confirm that W1243 rivals the low surface gravity features of even the youngest late-type L dwarfs confirmed in moving groups and we estimate an age of $\lesssim$ 150 Myr for our source. 

Given the discovery that W1243 was co-moving with a young K0 star, we also wanted to compare this source to known planetary mass companions. The bottom spectrum in the middle panel of Figure~\ref{fig:spectra-bands} shows the planetary mass companion 2MASSWJ 1207334-393254b (2M1207b; \citealt{Chauvin2004}) using the Sinfoni spectrum from \citet{Patience2010}. 2M1207b is classified as an L7 and is co-moving with an M8 dwarf in the $\sim$10 Myr TW Hya association. Compared to W1243, we find that 2M1207b has a sharper shape to the $H$ band but similar CO absorption in $K$ band.

In the bottom panel of Figure~\ref{fig:spectra-bands} we show a band by band comparison to the HR8799 planets (HR8799 bcde; \citealt{Marois2008,Marois2010}). We have pieced together the spectra of each source from Keck, Sphere, and/or GPI data, much of which was taken over multiple observing runs and extracted using different methods (\citealt{Barman2011,Barman2015}, \citealt{Greenbaum2018}, \citealt{Zurlo2016}). The HR8799 planets do not have confident spectral subtypes assigned but are classified as L type objects co-moving with an A5 star thought to be 40$\pm$5 Myr.    Compared to the HR8799 planets we find similarities and differences to each object depending on the band we examine.  The $J$ band data for both HR8799d and e match within uncertainties showing similar depths in $H_{2}$O absorption. The shape of the $H$ band for HR8799c matches that of W1243 as well.  HR8799b appears to have a much sharper triangular H-band shape than W1243 (also much sharper than any of its planetary siblings) and the data for HR8799e in both GPI and SPHERE hint at a similar sharp shape in that object.  The morphology of the $K$ band for each planet shows differing structure and we find the strongest similarities to HR8799c which appears to match most closely.  

From this comparison we can conclude that W1243 resembles both the isolated young L dwarfs as well as the young planetary mass companions with no obvious spectral differences that would differentiate it as having formed with one mechanism or another. 

\section{Evaluating the probability of Companionship for the System\label{sec:chancealign}} 
BD+60 1417 has a well measured parallax (22.2437 $\pm$ 0.0135 mas) and proper motion ([$\mu_{\alpha}$, $\mu_{\delta}$] = [-126.401$\pm$0.013, -64.141$\pm$0.015]\,mas\,yr$^{-1}$) in Gaia's eDR3 catalog (\citealt{eDR32020}). W1243 has proper motion components reported in the CatWISE2020 catalog ([$\mu_{\alpha}$,$\mu_{\delta}$] = [-133$\pm$8,-55$\pm$8]\,mas\,yr$^{-1}$; \citealt{Catwise2020}).  We estimate a photometric distance to W1243 using the $K_{s}$ band young L dwarf relation and a conservative ten percent uncertainty based on the parallax sample of late-type L dwarfs (44$\pm$4\,pc; \citealt{Faherty16}). We find that the $\mu_{\alpha}$ proper motion component and distance match well within 1$\sigma$ and the $\mu_{\delta}$ component matches just beyond 1$\sigma$.  We evaluated the probability of chance alignment in two ways.  

We examined the Gaia Catalog of Nearby Stars (GCNS; \citealt{GCNS}) -- which contains 331312 stars within 100pc -- and found there were 1602 objects with proper motion components that matched W1243 within 2$\sigma$ or 0.48\% of the catalog.  We then restricted that match to only sources that matched the photometric distance of W1243 within 20\% and were left with 159 objects or 0.05\% of the catalog.  Looking at the distribution of matches we found only one of those 159 objects -- BD+60 1417 at a 37$\arcsec$ physical separation-- was within 1 degree of W1243.  The next closest object was 1.5 degrees from our target making a co-moving kinematic match unreasonable.  \citet{GCNS} states that they are complete down to M9 dwarfs at 92\% confidence therefore this should be a robust examination of the spatial and kinematic distribution of nearby stars that might have matched W1243.

For a fully quantitative approach, we used a modified version of the BANYAN {$\Sigma$} code (\citealt{Gagne18}) called \texttt{common\_pm\_banyan} to evaluate co-evality.  This code -- described in \citet{BanyanComover} -- uses the sky position, proper motion, parallax and heliocentric radial velocity of a host star (with their respective measurement errors), and compares it to the observables of a potential companion (with their respective measurement errors) in order to determine a probability that the two stars are co-moving and thus gravitationally bound. When all kinematic observables are provided, a single spatial-kinematic model is built, consisting of a single 6-dimensional multivariate Gaussian in Galactic coordinates (XYZ) and space velocities (UVW). The observables of the potential companion are then compared to this model and the field-stars model of \citet{Gagne18} --a 10-component multivariate Gaussian model appropriate only within a few hundreds of parsecs of the Sun-- with Bayes' theorem by marginalizing over any missing kinematic observables of the companion star with the analytical integral solutions also detailed in \citet{Gagne18}. After inputting all values for both BD+60 1417 and W1243, the \texttt{common\_pm\_banyan} code yielded a 0\% probability of chance alignment for the system confirming all other approaches to determine its co-moving nature. Given our high level of confidence in the co-moving nature of this system we refer to the secondary as BD+60 1417B from here-on.

\begin{figure}[!ht]
\begin{center}$
\begin{array}{cc}

\includegraphics[width=3.1in]{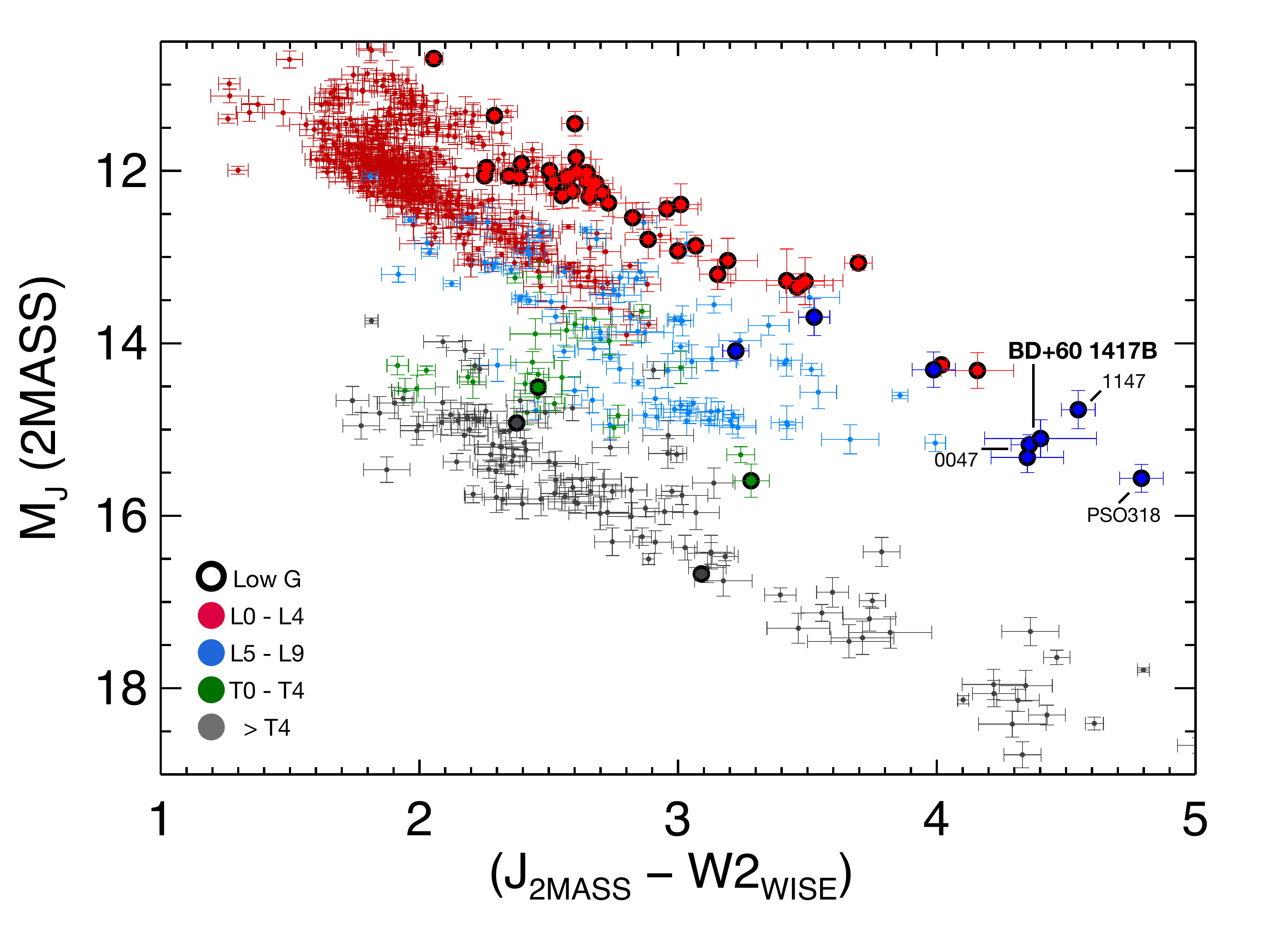}&\\
\includegraphics[width=3.1in]{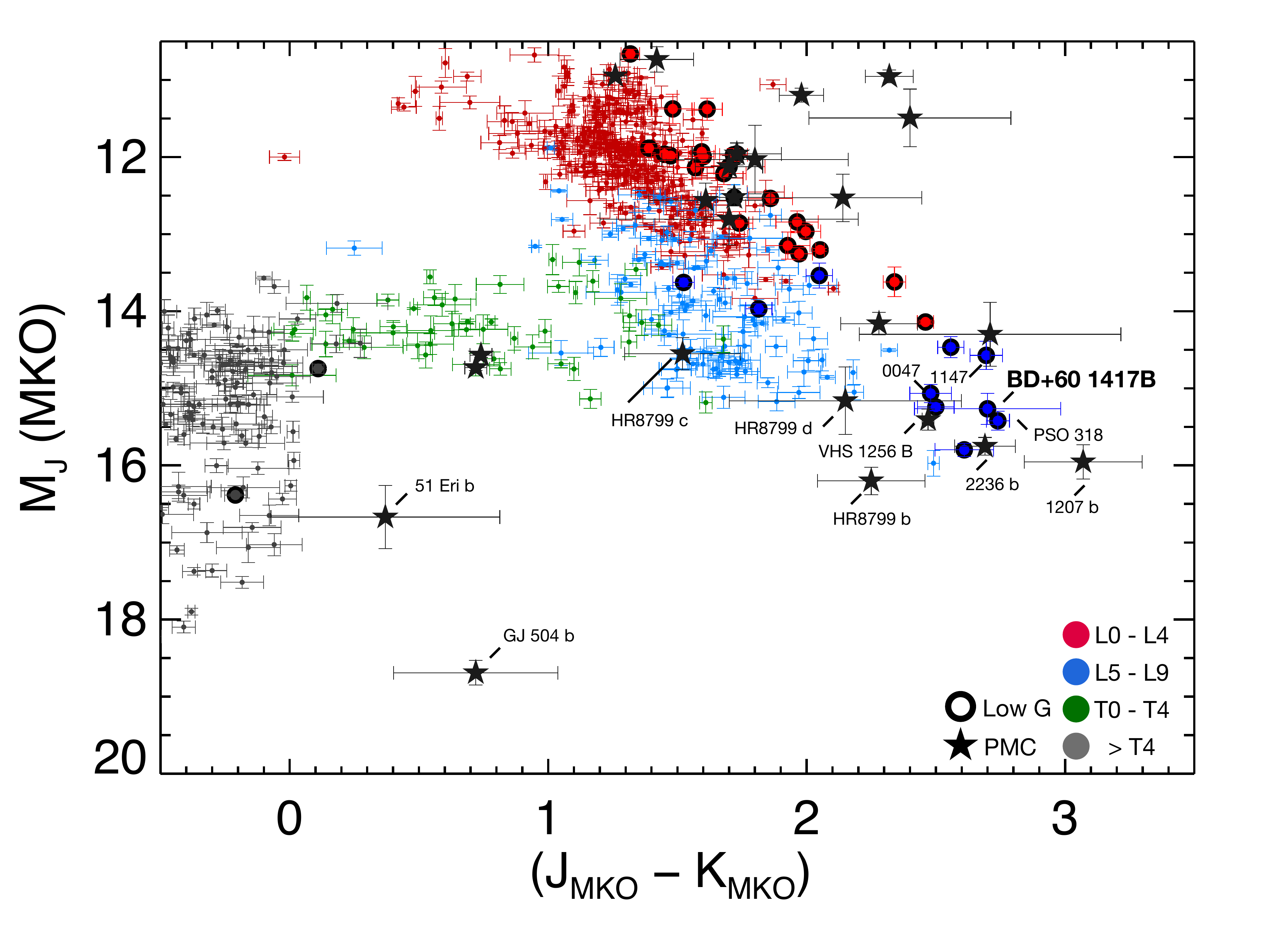} \\
\includegraphics[width=3.1in]{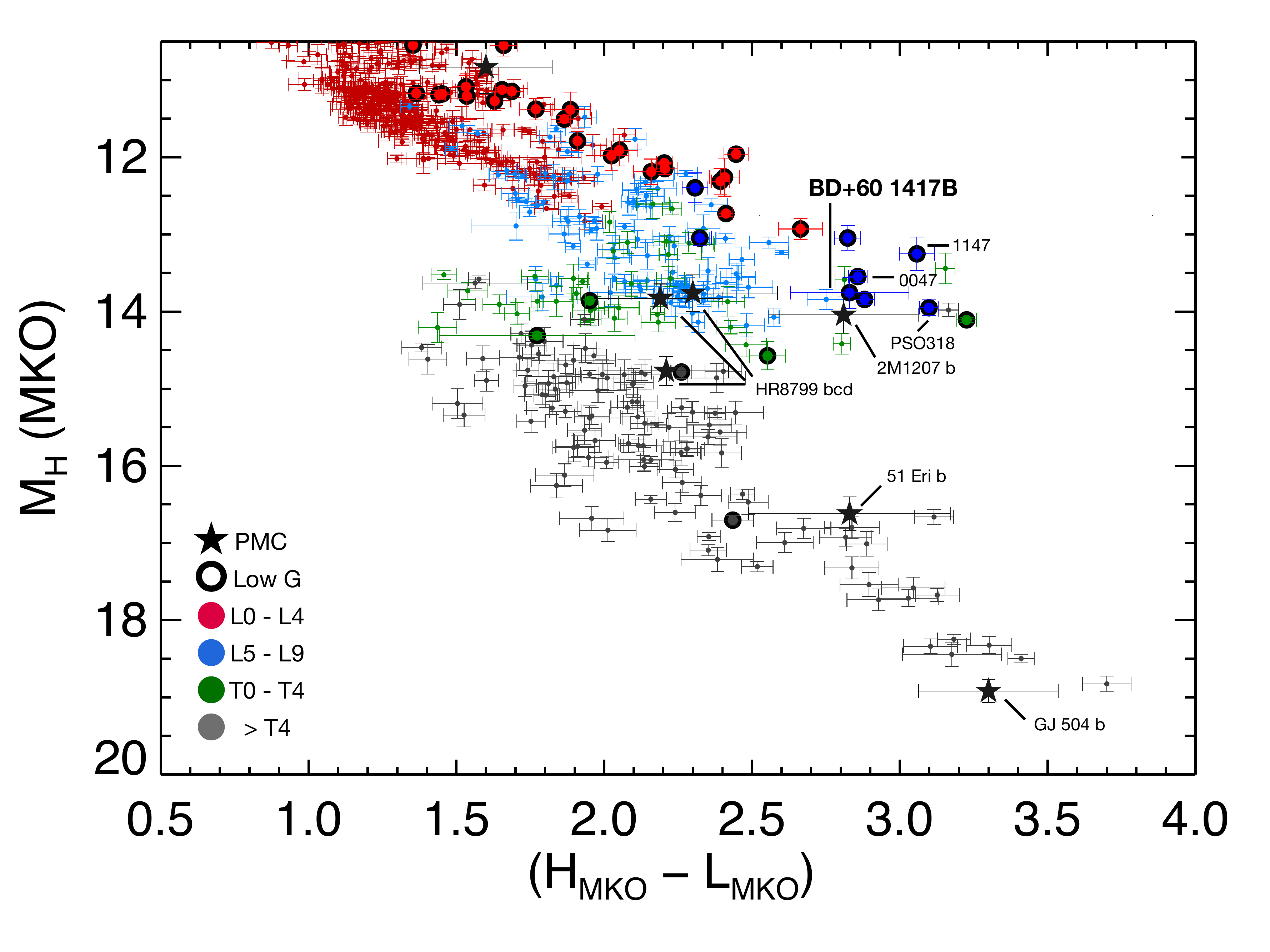} \\
\end{array}$
\end{center}
\caption{ The color-magnitude diagrams for field sources, low gravity brown dwarfs and directly imaged planetary mass companions color coded by spectral subtype. Top: M$_{J}$ (2MASS) vs. (J-W2) Middle: M$_{J}$ (MKO) vs. (J-K) (MKO) Bottom: M$_{H}$ (MKO) vs. (H-L) (MKO).  In the absence of L band photometry for brown dwarfs we used the relation in \citealt{Faherty16} to convert $WISE$ $W1$ photometry to L (MKO).}
\label{fig:CMD-Young}
\end{figure}

\begin{figure*}
\begin{center}
\includegraphics[width=0.8\linewidth]{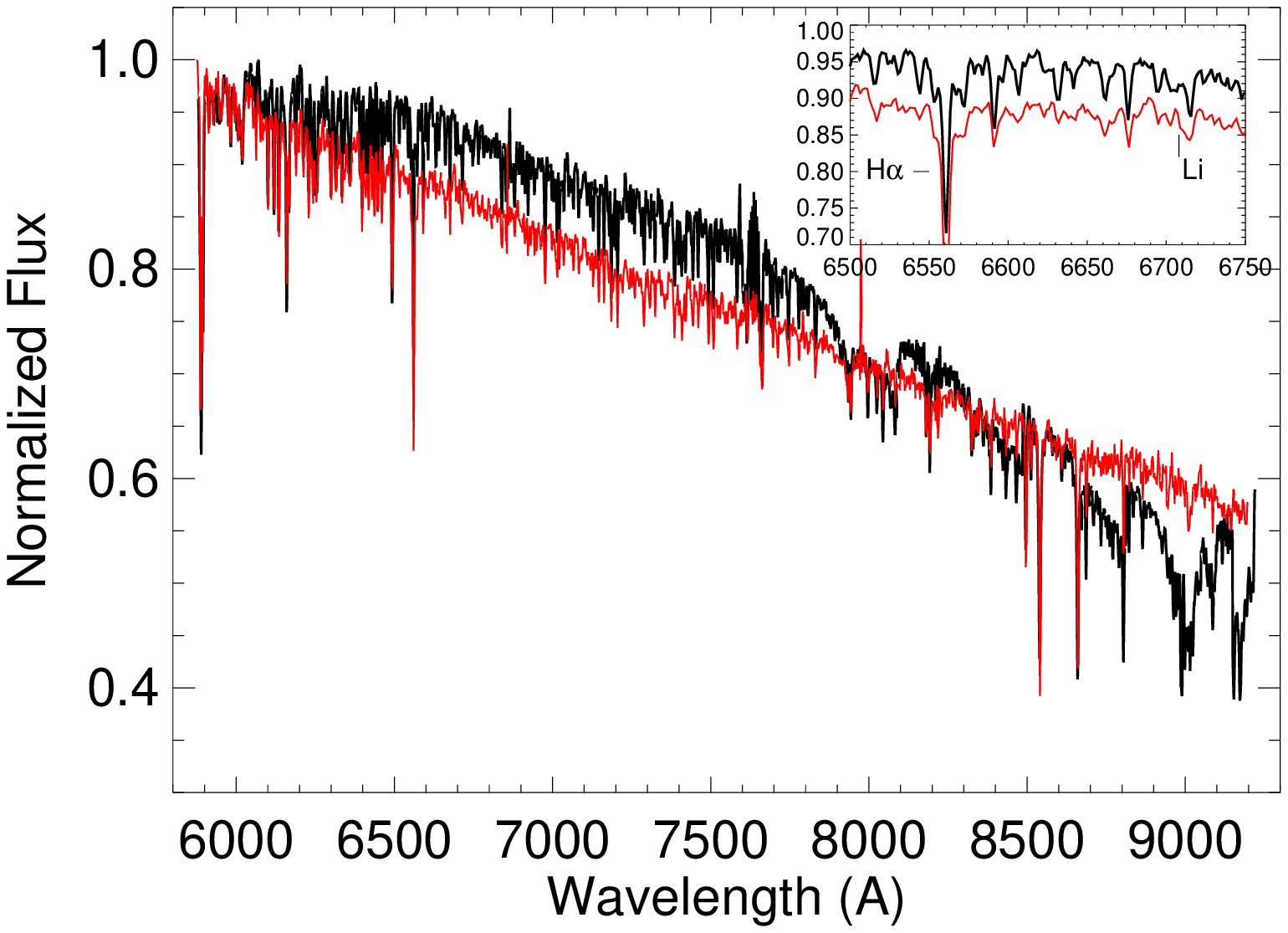}
\caption{The Kast optical spectrum of BD+60 1417 obtained for this work. In the inset we show the H$\alpha$ and Li features.  The flux is normalized over the peak of the full range.  Overplotted is a K0 star from the Sloan Digital Sky Survey in red.}
\label{fig:spectraprimary}
\end{center}
\end{figure*}

\section{BD+60 1417B on Color-Magnitude diagrams\label{sec:CMD}}
Using the parallax of the primary we can evaluate the position of BD+60 1417B on color-magnitude diagrams (CMDs) and compare/contrast it to other young L dwarfs and planetary mass companions.  The near- and mid-infrared CMDs have proven powerful diagnostic tools for deciphering the physical characteristics of young brown dwarfs (e.g. \citealt{Faherty16} and \citealt{Liu16}).  Recent studies have shown that low surface gravity brown dwarfs have redder near- to mid-infrared colors and fainter absolute magnitudes when compared to the sequence of field sources.  Young, warm, planetary mass companions share this observed feature, affirming that the two populations should be studied in tandem. 

Figure~\ref{fig:CMD-Young} shows three relevant panels of CMDs across near to mid infrared bands that are inclusive of the majority of planetary mass companions and isolated brown dwarfs with a parallax and relevant photometric measurements.  The (J-W2) CMD is inclusive of brown dwarfs only but allows us to put BD+60 1417B in context against all field and low gravity brown dwarfs.  Direct imaging studies of companions primarily observe with MKO $JHKL$ bands, so we also show the MKO (J-K) and MKO (H-L) CMDs. The majority of field and low gravity brown dwarfs have near infrared MKO band magnitudes but lack $L$ band photometry.  As such, in the (H-L) plot, we converted the W1 magnitude for field and low gravity brown dwarfs into L band magnitudes using the polynomial relation from \citet{Faherty16}.  Moreover, for BD+60 1417B we obtained synthetic MKO $J$ and $K$ photometry from the SpeX spectrum. The parallax sample was drawn from recent compilations by \citet{Best2020}, \citet{Kirkpatrick2021}, and \citet{Smart2019}. In each panel objects are color-coded by their spectral subtype, low surface gravity dwarfs (designated as $\gamma$ or VL-G only -- see \citealt{Faherty16}) are plotted as filled squares and planetary mass companions are plotted as filled five-point stars.  We collected planetary mass companion photometry from the compilation of \citet{Best2020}, as well as the NASA Exoplanet Archive\footnote{https://exoplanetarchive.ipac.caltech.edu/}. 

As can be seen in each plot, the low gravity brown dwarfs differentiate themselves as redder and simultaneously fainter than field objects of the same subclass.  As spectral subtypes increase across the L dwarfs, the low gravity objects deviate significantly from the field sample.  BD+60 1417B shares a similar position on the (J-W2) and (J-K) (MKO) CMDs as W0047 and PSO318.  As stated above, both sources are confirmed members of nearby moving groups with ages of $\sim$150 Myr and $\sim$24 Myr respectively.  Both sources have been noted as having particularly cloudy atmospheres with weather related variability (e.g. \citealt{Vos2018, Vos2019}, \citealt{Lew2016}).  BD+60 1417B also shares a spatial locus on the MKO (J-K) and (H-L) diagrams with planetary mass companions such as VHS1256B and 2M1207b.  VHS1256B was a recent target in a spectral monitoring campaign and was found to have strong near-infrared variability in HST/WFC3/G141 light curves (\citealt{Bowler2020}). \citet{Zhou2020} found the spectral variability of VHS1256B was consistent with predictions from partly cloudy models, suggesting that heterogeneous clouds are the dominant source of the observed modulations. The extreme positions of young, late-type objects on color-magnitude diagrams may be a strong signature of a particularly cloudy, active atmosphere (e.g. \citealt{Barman2011}, \citealt{Faherty12, Faherty16}, \citealt{Bowler2013}).  From the position of BD+60 1417B on each color-magnitude diagram we can conclude it is an excellent candidate for spectral and photometric variability studies.

\section{The Age of BD+60 1417\label{sec:age}}
As stated above, BD+60 1417 was categorized as a young star by \citet{Wichmann2003} and given an age range of 50 - 150 Myr.  In this section we revisit the parameters of the primary and re-evaluate the age using new indicators of youth from recent observations.  

\subsection{Age Indications from Li I absorption}
We obtained optical spectral data on the primary BD+60 1417 to investigate relevant molecular features.  Figure ~\ref{fig:spectraprimary} shows the 6000 - 9000\AA~spectrum with an inset of the wavelength regime for H$\alpha$ and Li I.  For comparison, we have overplotted the K0 star SDSS J012703.04+383611.1 that was spectroscopically characterized with the Sloan Digital Sky Survey.  There looks to be a slight slope in our Kast spectrum compared to the SDSS data, potentially due to imperfections in the reduction. However we find comparable features confirming the K0 spectral subtype. In \citet{Wichmann2003}, the authors reported a Li I measurement corrected for Fe I of 96 m\AA.  We can not confirm that value with our data given our resolution of $\sim$6000\AA~means the feature is too contaminated by Fe I.   Using the measured value of Li I absorption from \citet{Wichmann2003} we can compare to values for other young stars near the Sun.  Similar to the conclusions of \citet{Wichmann2003} we find that BD+60 1417 is consistent with values for the Pleiades indicating that this source is likely of a similar age range ($\sim$110 - 125 Myr).

\subsection{Age Indications from Gyrochronology}
\begin{figure}[ht!]
\begin{center}
\includegraphics[width=0.5\linewidth]{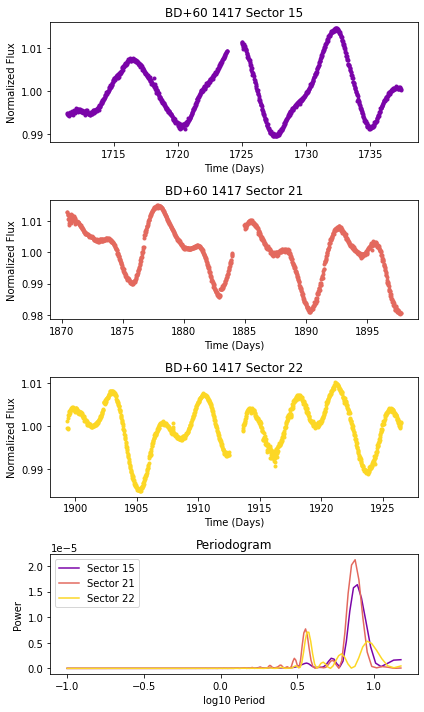}
\caption{The TESS light curve in sectors 15, 21, and 22 as well as the power spectrum of the Lomb Scargle periodogram for these data.}
\label{fig:lightcurves_ls}
\end{center}
\end{figure}

BD+60 1417 was observed by NASA's Transiting Exoplanet Survey Satellite (TESS; \citealt{TESS}) telescope in sectors 15, 21, and 22 in both 30 minute and 2 minute cadences. From the full frame images we generated light curves for each sector using simple aperture photometry and background subtraction. We present these light curves in Figure~\ref{fig:lightcurves_ls}, along with the results of running each through a Lomb-Scargle periodogram (\citealt{lomb_1976}, \citealt{scargle_1982}). The periodogram shows a strong peak at around 7.5 days in both sector 15 and 21, and at the half harmonic of 3.7 days in sector 21 and 22. The morphology of the light curve changes between sectors; in sector 15 the shape is a simple sinuosoid, while in sectors 21 and 22 it is a double dip pattern. Between sectors 21 and 22 the shape of the double dip changes between cycles. Likely we are seeing the evolution of sunspots across the surface of the star.  Averaging across sectors we report a rotation period of 7.50 $\pm$ 0.86 days.

\begin{figure}[ht!]
\begin{center}
\includegraphics[width=0.8\linewidth]{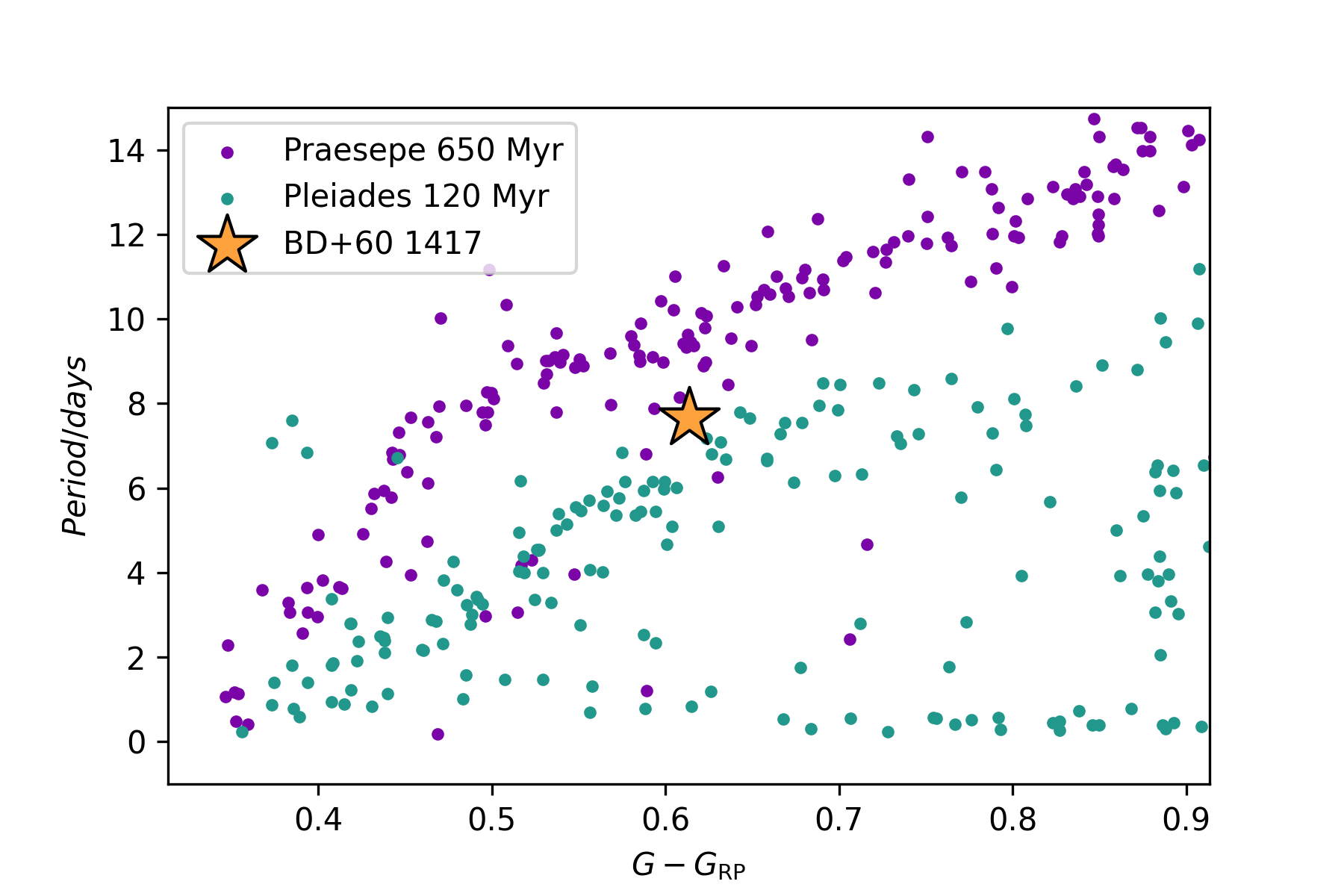}
\caption{The color-period plot for BD+60 1417 (five point star) in context with the 650 Myr Praesepe (purple filled circles) and 120 Myr Pleiades (blue filled circles) clusters. }
\label{fig:color_period}
\end{center}
\end{figure}

We put BD+60 1417 on a color-period plot in Figure~\ref{fig:color_period}. To provide context we include rotation periods from the gyrochronology benchmark clusters the Pleiades (\citealt{rebull_2016_pleiades}) and Praesepe (\citealt{douglas2017}) at $\sim$120 Myr and 650 Myr respectively. For its color BD+60 1417 is rotating faster than the majority of Praesepe objects, and slightly slower than the majority of Pleiades objects.  We also report an average amplitude variation across the three sectors of 2.23\%. While variation for a single star can depend on various factors such as inclination and activity cycle, this is consistent with the amplitude variation of a $\sim$100 Myr object (e.g. \citealt{morris_2020}). Conservatively, we find that a rotation period analysis indicates an age range of 100-650 Myr for BD+60 1417. 

\subsection{Age Indications from Color-Color and Color-Magnitude Diagrams}
\begin{figure}[!ht]
\begin{center}$
\begin{array}{cc}

\includegraphics[width=0.6\linewidth]{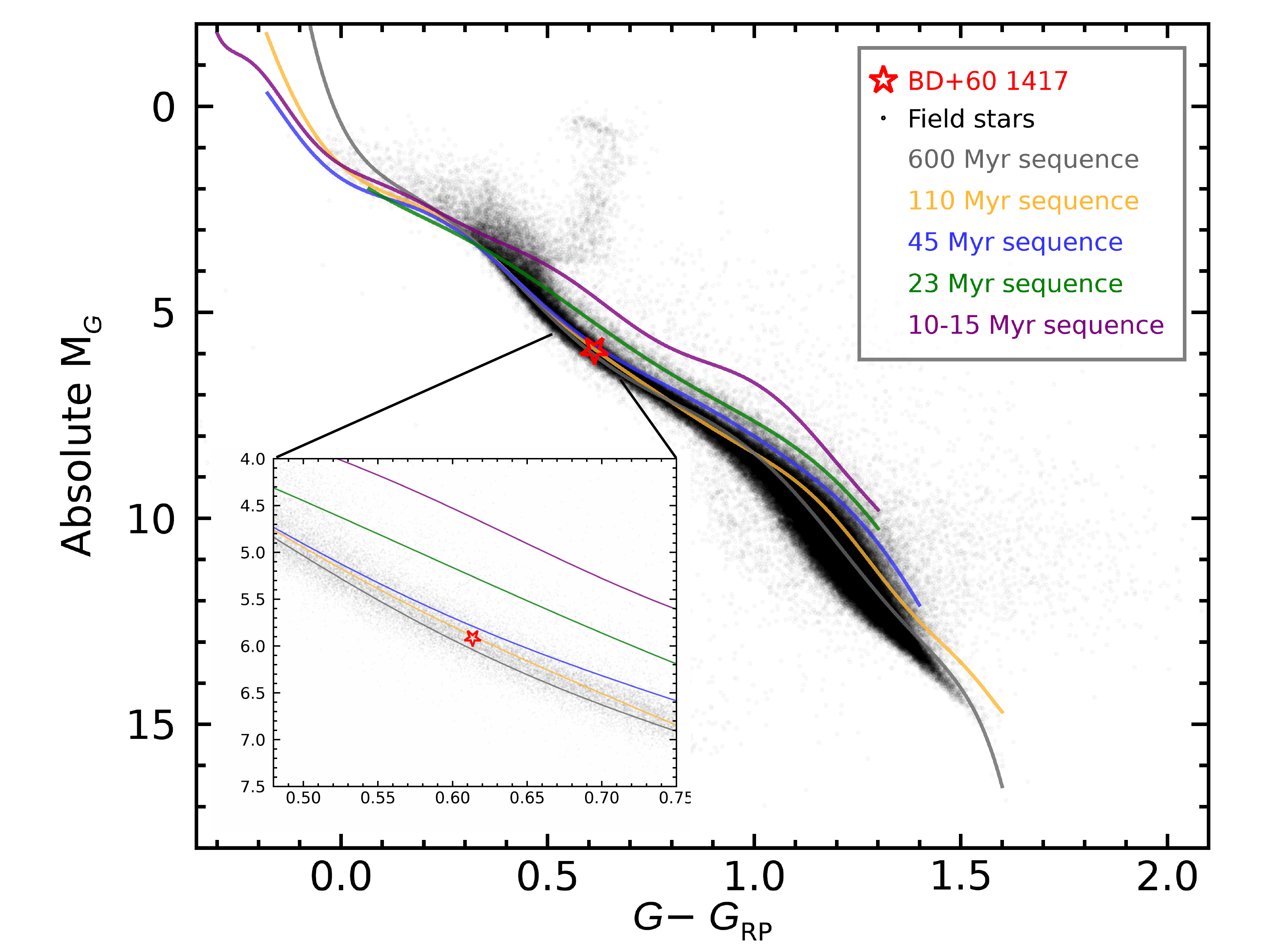}\\
\includegraphics[width=0.6\linewidth]{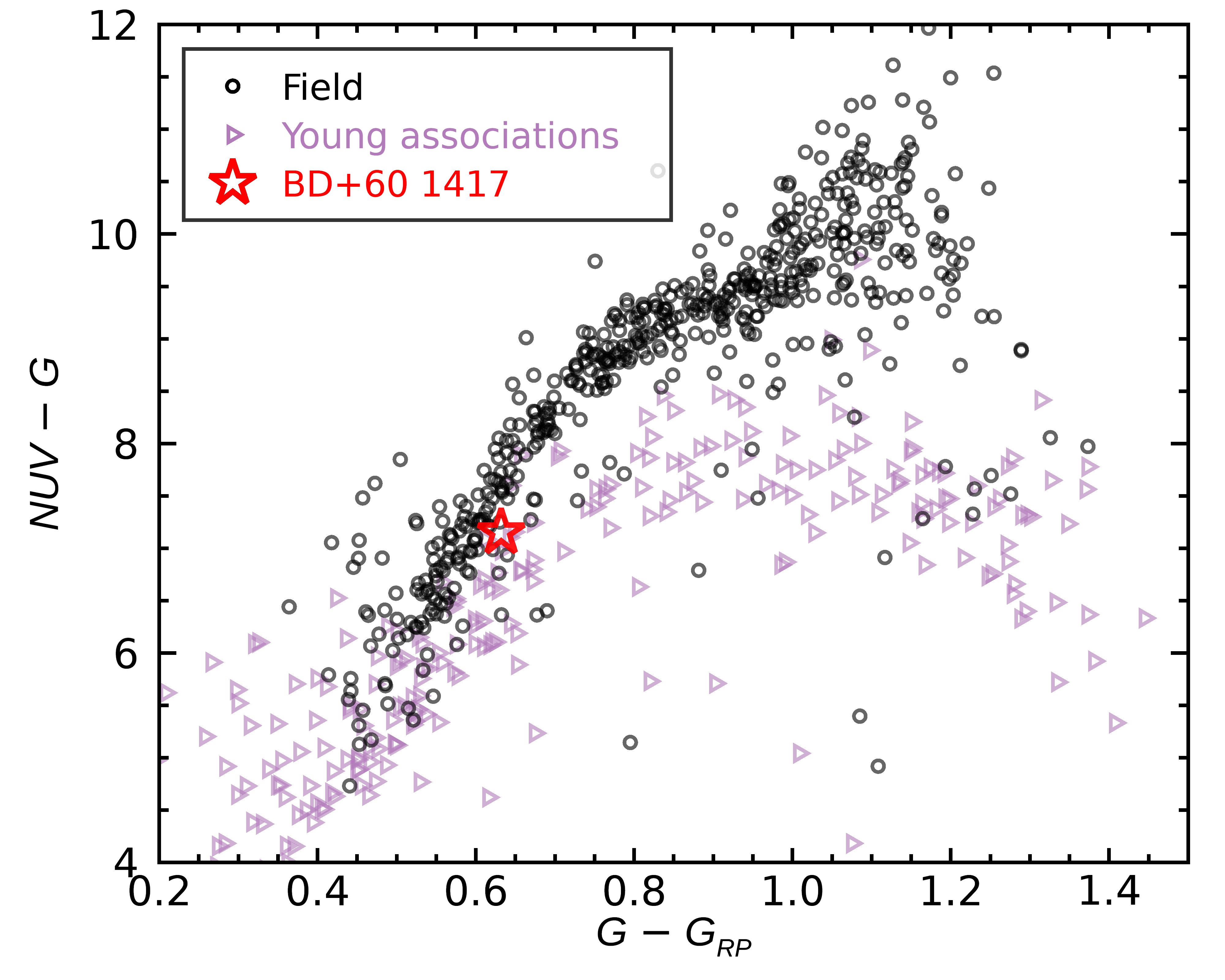}\\
\end{array}$
\end{center}
\caption{ {\bf Top:} The Gaia color-magnitude diagram for both the field sample and the 150pc sample of young stars discussed in \citet{GagneFaherty2018}. We have zoomed in on the position of BD+60 1417 as an inset to show the closest isochrone. {\bf Bottom:} GALEX to Gaia (NUV-G) color vs Gaia (G-$G_{RP}$) for both the field sample and the 150pc sample of 10-150 Myr young stars from \citet{GagneFaherty2018}.  The position of BD+60 1417 is marked by a red five-point star.}
\label{fig:CMDandCC}
\end{figure}

Pulling together all of the catalog detections of BD+60 1417 we can look at where it stands on color-color and color-magnitude diagrams in order to re-assess the extent of its youth.  Using the Gaia parallax and photometry we can compare BD+60 1417 to the 150pc sample of young stars from \citet{GagneFaherty2018}.  The top panel of Figure~\ref{fig:CMDandCC} shows the empirical isochrones from \citet{GagneFaherty2018} for objects across the color-magnitude diagram at ages of 10-15 Myr, 23 Myr, 45 Myr, 110 Myr, and 600 Myr with a zoom-in on the position of BD+60 1417 as an inset.  As one goes from younger to older ages for the main sequence, the isochrones shift to fainter absolute magnitudes.  BD+60 1417 is situated at a CMD position where the 45 Myr, 110 Myr, and 600 Myr isochrones are difficult to distinguish.  We can conclude that it does not share properties with 23Myr 
sources and the inset shows that it is most consistent with the 110 Myr sequence.

Combining the ultraviolet flux detected from BD+60 1417 from the Galaxy Evolution Explorer (GALEX) survey we can use the (NUV-G) color as a proxy for enhanced magnetic activity which has also been shown to scale with younger ages (e.g. \citealt{Gagne2020}, \citealt{GagneVolans}).  The bottom panel of Figure~\ref{fig:CMDandCC} shows the field sequence as black unfilled circles and the young sample (10 - 150 Myr group members) as purple upward facing triangles.  BD+60 1417 falls just outside the tighter field sequence.  At this temperature (or (G-G$_{RP}$) color) the NUV is not very diagnostic of youth however it is safe to say that the position of BD+60 1417 is consistent with a 10-150 Myr source.

Similarly, we can compare the X-ray luminosity for our source as calculated using the ROSAT detection (\citealt{Rosat}) to the sample of young stars near the Sun.   With a log (L$_{x}$)$\sim$ 29 (erg s$^{-1}$) BD+60 1417 is consistent with a star of age $\sim$150 Myr or younger (see for example \citealt{Kastner2003}, \citealt{Malo2014}, \citealt{Gagne2020}).  

Putting all these primary star diagnostics together with our assesment of the secondary, we decide to maintain the 50 - 150 Myr age range given by \citet{Wichmann2003} as our re-analysis is consistent with this value.

\begin{figure*}
\begin{center}
\includegraphics[width=1.0\linewidth]{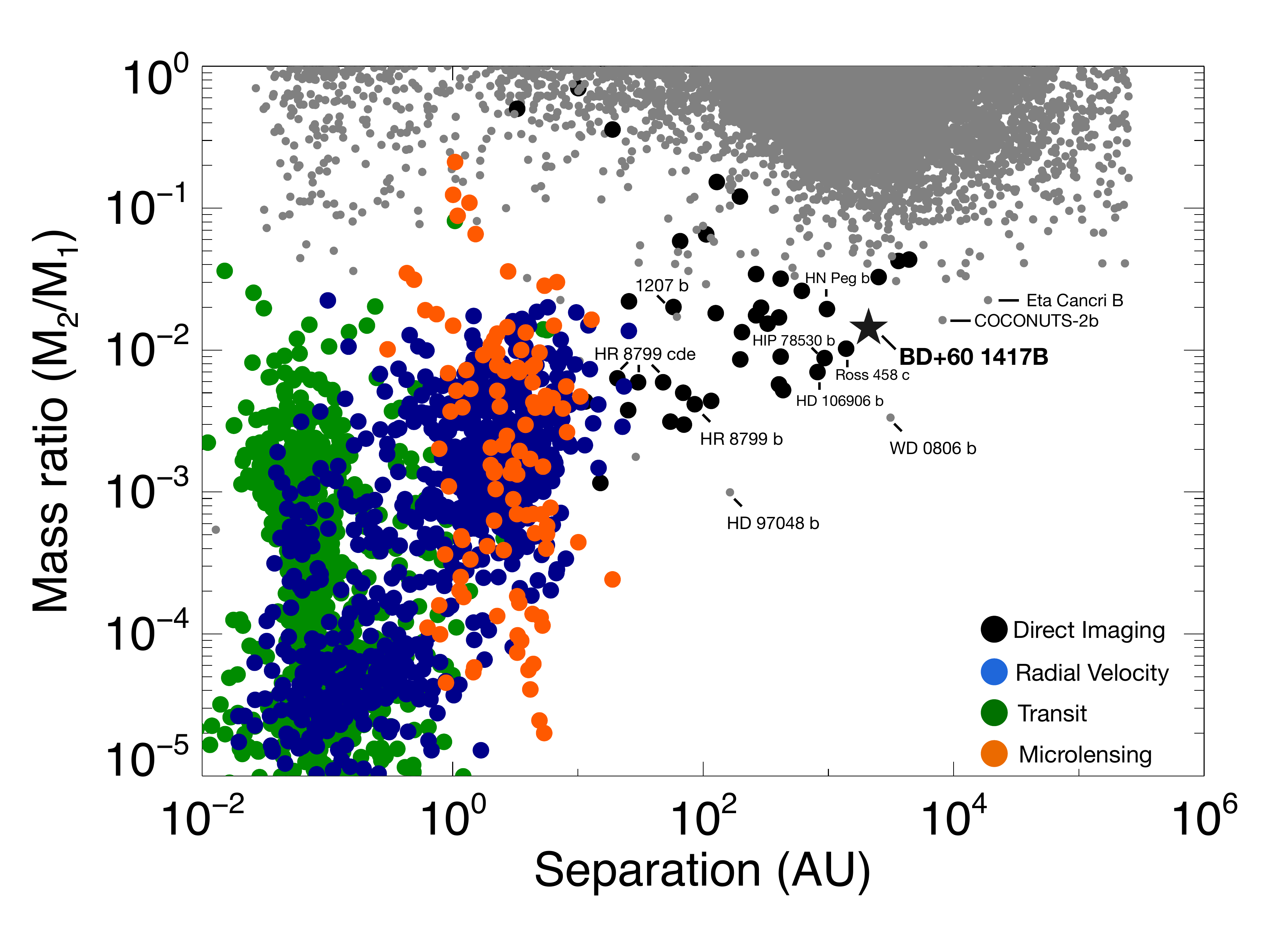}
\caption{Mass ratio versus separation for a collection of companions from stellar to planetary.  We have color coded the exoplanet detections (as defined by the exoplanet archive) by the detection method of the source.}
\label{fig:Mass-Sep}
\end{center}
\end{figure*}

\section{Fundamental Parameters\label{sec:fundamentals}}
In order to evaluate the physical characteristics of the BD+60 1417 system, we needed to obtain fundamental parameters of each component, such as mass, luminosity, radius, etc.  

For the primary, we examined the range of values cited for BD+60 1417 in the literature (T$_{eff}$ for instance is listed as 5142 K in \citealt{Starhorse2018}, 4981 K in \citealt{GaiaDR2}, 5198 K in \citealt{Miller2015}, and 4993$\pm$124 K in \citealt{TIC2018}).  There is clear variety in values listed.  For the analysis that follows we chose to use the fundamental parameters listed in the TESS Input Catalog (TIC; \citealt{TIC2018}) because they present mass, radius, log(g), L$_{bol}$ uniformly with uncertainties while other catalogs limit to just a few parameters.   We note that the TIC does not necessarily take into account the youth of this source and therefore the uncertainties and the values such as the radius may be underestimated.  Given that we wanted to understand the mass ratio of the system in context with other wide co-movers, we choose to use the mass published in  \citet{Metchev2009} which did take into account the youth of the source.  All parameters for BD+60 1417 can be found in Table~\ref{tab:param}.  

We used the $v\sin(i)$, radius, and rotation rate for BD+60 1417 to infer its viewing inclination.  Assuming rigid body rotation, the $\sin(i)$ distribution can be calculated by:
\begin{equation}
    \sin(i) = \frac{P v\sin(i)}{2 \pi R}
\end{equation}
where $P$ is the rotation period, and $R$ is the stellar radius. We use Monte Carlo analysis to determine the $\sin (i)$ and inclination distributions, using Gaussian distributions for $v \sin (i)$, period and radius. Monte Carlo analysis is especially well-suited to this computation as it can take into account a variety of distributions for $v\sin(i)$, $P$ and $R$ \citep[e.g.][]{Vos2017, Vos2020}, although we assume Gaussian distributions in our case.   \citet{Wichmann2003} report the $v\sin (i)$ of BD+60 1417 as $11$\,km\,s$^{-1}$, citing an approximate error for their entire sample of 10\%. However, measuring low $v \sin (i)$ values is particularly difficult due to the weak rotational broadening of the spectral lines for slow rotators. When values for $\sin (i)$ fell above 1, we set these values to 1 as discarding them would bias our results towards lower inclinations \citep{Vos2017}.
Since $11$\,km\,s$^{-1}$ is one of the lowest $v \sin (i)$ values reported in the Wichmann study, we assume larger error bars of $3$\,km\,s$^{-1}$. Using the parameters listed in Table~\ref{tab:param}, we find that the  majority of the $\sin (i)$ distribution ($\sim96\%$) falls above 1, with a median $\sin(i)=2$. This results in an inclination distribution with a sharp peak at $90^{\circ}$, or equator-on. A doubling of the radius or a halving of the rotation period would bring the median value to 1. Although we observe a period harmonic at half our reported rotation period, we consider it unlikely to be the true period due to the repeating double-peaked pattern observed  in Figure~\ref{fig:lightcurves_ls}. We consider it more likely that the radius is inflated due to its youth (e.g., \citealt{Somers2015, Somers2017}). Either way, the measured rotation rate and $v\sin(i)$ values and estimated radius suggest that  BD+60 1417 is inclined close to equator-on.

As shown in Figure~\ref{fig:lightcurves_ls}, the light curve of BD+60 1417 shows some starspot evolution but is otherwise a clean rotation curve.  We find no signature of a transiting world over the TESS cadences even though we are staring at the star across the equator.

For the secondary, to determine all fundamental parameters we followed the prescription of \citet{Filippazzo15}. We used the Gaia eDR3 parallax of BD+60 1417 combined with all available photometric and spectroscopic information on BD+60 1417B to produce a distance calibrated spectral energy distribution.  By integrating over the SED, we directly calculated the bolometric luminosity. Using the evolutionary models of \citet{Saumon08} paired with the 50 - 150 Myr age range cited above, we obtained a radius range and semi-empirically obtained estimates for the $T_{\rm eff}$, mass, and log(g).  All fundamental parameters are listed in Table~\ref{tab:param}.

\section{Characteristics of the BD+60 1417 System\label{sec:characteristics}}
Figure~\ref{fig:Mass-Sep} shows the separation versus mass ratio of a wide variety of companion systems.  For stellar binaries, we drew from the literature compilations in \citet{Faherty10} appended with recent discoveries of 50pc co-moving companions in \citet{ElBadry2021}.  For brown dwarf companions we used the recent compilation by \citet{Faherty2020} appended with a thorough literature search performed for this work of MLT companions in systems with total mass $>$ 100 M$_{Jup}$ (for example adding recent discoveries by \citealt{Zhang2021B} and Schneider et al. 2021) and for planetary companions we used the NASA exoplanet archive database.   

From the bottom left corner of Figure~\ref{fig:Mass-Sep} to the top right corner we are sampling objects that form via traditional planetary processes (e.g. core accretion) to objects that form via traditional stellar processes (e.g. cloud fragmentation).  As we pass from the lower left to the upper right, we may sample a mix of those formation processes or a clear delineation between differing populations (e.g. \citealt{Kratter2010}).  We also must note that there are observing biases on this plot given that transiting planets and radial velocity planets have limitations to the maximum separation they can be detected given the long baseline required to find more distant planets.  

We have color coded Figure~\ref{fig:Mass-Sep} by exoplanet detection method as the question at hand is whether BD+60 1417B should be considered a young companion formed via binary star processes or a young planet formed via accretion or similar processes.  Directly imaged companions occupy a sparsely sampled region of Figure~\ref{fig:Mass-Sep} which is detached from the locus of stellar binaries (grey overdensity at upper right) and exoplanet companions (lower left).  The BD+60 1417 system sits squarely between the two loci in a very similar position to the Ross 458, HIP 78530, HN Peg, and HD 106906 systems.  It has the largest separation known for its mass ratio\footnote{The estimated mass ratio for BD+60 1417 is just smaller than that of the recently published COCONUTS-2b source by \citet{Zhang2021B}}, making it a boundary setting object on Figure~\ref{fig:Mass-Sep}.  

Assuming the mass of the primary is $\sim$1.0 M$_{\sun}$ (from \citealt{Metchev2009}), the secondary is 15 M$_{Jup}$, and the physical separation is 1662 AU x 1.26 (for orbital inclination angle correction; see \citealt{Fischer1992}), we find a binding energy of $\sim$1.204 x 10$^{41}$ erg.  In our analysis of main sequence star co-moving systems, brown dwarf and exoplanet companions we find this source in the lowest 4\% in terms of binding energy. We also find it joins just a handful of companion objects (GU Pscb, USco 1621b, USco 1556b, Ross 458c, FU Taub, COCONUTS-2b, 2MASSJ21265040-8140293b, WD0806b, and Ross 19B) with estimated masses $<$ 20 M$_{Jup}$ and a separation $>$ 1000 AU\footnote{The separation is evaluated as 1.26x the projected separation to account for inclination angle}.  Among those systems and with the exception of WD0806 which has evolved off the main sequence, BD+60 1417 has nearly twice the primary mass of any other.

\section{Discussion\label{sec:discussion}}
Determining the formation mechanism for planetary mass objects (companion or isolated) remains difficult. It is very likely that as we move down in mass, there are contributions to the mass function from competing mechanisms that can create identical looking objects.  BD+60 1417B is an object with photometric and spectral characteristics perfectly in line with isolated and companion $<$ 20 M$_{Jup}$ objects.  It does not differentiate itself from closer planets in any of its fundamental parameters.  In fact, it resembles those sources so well it should be considered an important laboratory for investigating weather related atmospheric phenomenon relevant to giant planet analyses. 

 \citet{Kirkpatrick2021} recently analyzed the mass function of the 20pc sample of brown dwarfs into the Y dwarf regime and discovered that the current sample implies that the cutoff of star formation is lower than 10 M$_{Jup}$ with some hint from moving group members that it might be lower than 5 M$_{Jup}$.  Based on these cut-off values, even the lower mass limit for BD+60 1417B would argue for its formation mechanism still aligned with that of higher mass isolated brown dwarfs. However, the BD+60 1417 system occupies a unique and sparsely populated region in Figure~\ref{fig:Mass-Sep} and BD+601417B has a mass above the 5 M$_{Jup}$ cut-off. The dearth of brown dwarf companions close-in to their host stars is a well known phenomenon and has been dubbed ``the brown dwarf desert".  Studies such as \citet{Raghavan2010}, \citet{Metchev2009}, \citet{Brandt2014} conducted detailed investigations on the companion frequency for low mass secondaries.  \citet{Metchev08} concluded that the companion mass function follows the same universal form over the entire range between 0 - 1590 AU in orbital semi-major axis and $\sim$0.01–20 $M_{\sun}$ in companion mass therefore the brown dwarf desert is a logical extension of the field mass function.  \citet{Brandt2014} concluded that many of the directly imaged exoplanets known formed by fragmentation in a cloud or disk, and represent the low-mass tail of the brown dwarfs. It is completely logical that BD+601417B formed as a binary system with BD+60 1417 and is simply a rare, planetary-mass product of the lowest mass limit of star formation.

Even still, there remains the possibility that BD+60 1417B formed closer in through core accretion or disk instability and was dynamically disturbed to its current position.  Recent investigations of the planetary mass companion HD 106906b (estimated mass$\sim$11$\pm$2 M$_{Jup}$ at 738 AU) suggested that a stellar fly-by may have disturbed the companion to its current position and orbit given that the planet is not coplanar with the surrounding disk (\citealt{DeRosa2019}).  Other works have shown the significant influence that stellar fly-bys may have on young forming planetary systems (e.g. \citealt{Malmberg2011}, \citealt{Picogna2014}) and the potential for disturbing or even ejecting a companion.  While there is currently no evidence for a dynamic disturbance to this system, the possibility can not be ruled out.   

 \citet{Bowler2020b} used eccentricity values for low mass companions to determine that when subdivided by companion mass and mass ratio, the underlying distributions for
giant planets and brown dwarfs show significant differences. The large separation of this system and consequent significant period to complete an orbit precludes any eccentricity measurement to place it in context.  However, another avenue of investigation could be using the inclination angle for BD+60 1417B and compare that to BD+60 1417.  As stated in section~\ref{sec:fundamentals}, BD+60 1417 appears to be viewed equator on.  A measurement of the rotation rate and vsini for BD+60 1417B would allow us to calculate the viewing angle (e.g. \citealt{Vos2017}) and compare it to that of the primary.  

Yet another avenue that might help disentangle the formation mechanism for this companion is a comparison of the bulk composition of the secondary with the primary.  The spectral inversion technique can successfully recover gas abundances and atmosphere properties of brown dwarfs (e.g. \citealt{Burningham2017, Burningham2021}, \citealt{Line2017}) including C/O, Mg/Si, etc. Previous studies of giant planetary mass companions (e.g. \citealt{Konopacky2013}) have used the C/O ratio comparison between components to speculate on a formation route.  Recent work by \citet{Gonzales2020} used the spectral inversion technique to confirm two brown dwarfs as common origin based on similar composition measurements including a measurement of C/O for both components.  The BD+60 1417 system is an excellent candidate for a spectral inversion approach to determine if chemical composition could help differentiate the formation mechanism.  While that is outside of the scope of this work, the data collected here-in can and should be used by retrieval codes such as BREWSTER (\citealt{Burningham2017}, \citealt{Burningham2021}) which uses complex cloud approaches to examine the detailed atmospheric chemistry of a given source.

\section{Conclusions\label{sec:conclusions}}
In this work we have discussed the discovery of a young L dwarf companion to a nearby young K0 primary through the Backyard Worlds: Planet 9 citizen science project.  Using SpeX prism follow-up spectroscopy, we classify BD+60 1417B as an L6-L8$\gamma$ and find it to be one of the reddest known L dwarfs in the near infrared.  We find its spectral morphology is similar to other 10 - 150 Myr L dwarfs such as W0047, PSO 318, and W2244, and --similar to those sources-- its shape suggests that it is a cloudy object with the potential to demonstrate cloud-driven variability. The position on near- and mid-infrared color-magnitude diagrams -- using the parallax of the primary -- places BD+60 1417B as redder and fainter than the field sample, another hallmark characteristic of $<$150 Myr L dwarfs.  BD+60 1417B shares spectral and photometric properties with both isolated low gravity objects and young giant planets orbiting higher mass stars (e.g. HR8799bcde).  Using the parallax and spectral data we compute a spectral energy distribution and find log($L_{\rm bol}$/$L_{\odot}$)=-4.33$\pm$0.09, T$_{eff}$=1303$\pm$74 K, and Mass=15$\pm$5M$_{Jup}$ for BD+60 1417B.

The angular and physical separation of this system are 37$\arcsec$ and 1662 AU respectively.  We investigated the mass ratio vs. separation for the BD+60 1417 system and found it occupied a unique and sparsely populated region of the diagram between two loci of binary stars and planet companions.  It belongs to a sub-population of only 9 objects with an estimated mass $<$20 M$_{Jup}$ that orbit their host star at more than 1000AU. Among that small list it has a primary mass almost twice that of any other main sequence system (the exception being WD0806 which has a white dwarf host star with a progenitor mass of $\sim$2M$_{sun}$). We can not conclude a formation mechanism from any of the parameters presented in this paper (core accertion, disk instability, cloud fragmentation) but we are optimistic that future retrieval or viewing angle comparisons with the primary will shed light on how this system formed.  BD+60 1417B is an exciting target for future missions such as JWST which will open the mid-infrared spectral window on studying the clouds and composition of the source.

\begin{acknowledgments}
The Backyard Worlds: Planet 9 team would like to thank the many Zooniverse volunteers who have participated in this project, from providing feedback during the beta review stage to classifying flipbooks to contributing to the discussions on TALK. We would also like to thank the Zooniverse web development team for their work creating and maintaining the Zooniverse platform and the Project Builder tools. This research was supported by NASA Astrophysics Data Analysis Program grant NNH17AE75I as well as NASA grant 2017-ADAP17-0067, and National Science Foundation Grant No.'s 2007068, 2009136, and 2009177. F.M.~also acknowledges support from grant 80NSSC20K0452 under the NASA Astrophysics Data Analysis Program.  E.G. and J.F. acknowledge support from the Heising-Simons Foundation. This research has made use of: the Washington Double Star Catalog maintained at the U.S. Naval Observatory; the SIMBAD database and VizieR catalog access tool, operated at the Centre de Donnees astronomiques de Strasbourg, France (\citealt{Ochsenbein00}); data products from the Two Micron All Sky Survey (2MASS; \citealt{Skrutskie06}), which is a joint project of the University of Massachusetts and the Infrared Processing and Analysis Center (IPAC)/California Institute of Technology (Caltech), funded by the National Aeronautics and Space Administration (NASA) and the National Science Foundation; data products from the Wide-field Infrared Survey Explorer ($WISE$; and \citealt{Wright10}), which is a joint project of the University of California, Los Angeles, and the Jet Propulsion Laboratory (JPL)/Caltech, funded by NASA.  This work has made use of data from the European Space Agency(ESA) mission Gaia, processed by the Gaia Data Processing and Analysis Consortium. Funding for the DPAC has been provided by national institutions, in particular the institutions participating in the Gaia Multilateral Agreement. This research has made use of the NASA Exoplanet Archive, which is operated by the California Institute of Technology, under contract with the National Aeronautics and Space Administration under the Exoplanet Exploration Program.   
\end{acknowledgments}

\startlongtable
\begin{deluxetable*}{cccccc}
\tablecaption{Measured Parameters\label{tab:param}}
\tablehead{
\colhead{Parameter} & \colhead{BD+60 1417} & \colhead{BD+60 1417B} & \colhead{System} & \colhead{Units} & \colhead{Reference}}
\colnumbers
\startdata
{\bf ASTROMETRY}\\
\hline
\hline
$\alpha$ &  190.88751026191\tablenotemark{a}& 190.8838648 &$\cdots$ & deg &1,5\\
$\delta$ &  +60.01435471736\tablenotemark{a}& 60.0239567  &$\cdots$ & deg &  1,5\\
$\varpi$ & 22.2437\,$\pm$\,0.0135 &$\cdots$    &$\cdots$ & mas &1\\
Distance       & 44.957$\pm$0.027\tablenotemark{b} & 44$\pm$4\tablenotemark{c} &  & pc & 2\\
$\mu_{\alpha}$ & -126.401\,$\pm$\,0.013 & -133\,$\pm$\,8 & $\cdots$ & mas\,yr$^{-1}$ & 1,52\\
$\mu_{\delta}$ & -64.141\,$\pm$\,0.015  & -55\,$\pm$\,8 & $\cdots$ & mas\,yr$^{-1}$ & 1,5\\
\hline
\hline
{\bf ROTATION}\\
\hline
\hline
Prot     & 7.50$\pm$0.86 &$\cdots$&$\cdots$& days & 2\\
vsini &$11\pm3$&$\cdots$&$\cdots$&km\,s$^{-1}$&8\\
inc angle  & {$\sim90^{\circ}$} &$\cdots$&$\cdots$& deg & 2\\
\hline
\hline
{\bf PHOTOMETRY}\\
\hline
\hline
$G_{BP}$ & 9.652961$\pm$0.003343 & $\cdots$ & $\cdots$ & mag &  1\\
$G$     & 9.183514$\pm$0.002805 & $\cdots$ & $\cdots$ & mag &  1\\
$G_{RP}$ & 8.551546$\pm$0.003985 & $\cdots$ & $\cdots$ & mag &  1\\
Pan-STARRS $y$ & $\cdots$ & 20.481$\pm$0.178& $\cdots$ & mag &  1\\
2MASS $J$ & 7.823$\pm$0.020  & $>$17.452 & $\cdots$ &  mag & 3\\
2MASS $H$ & 7.358$\pm$0.016  & $>$16.745 & $\cdots$ &  mag & 3\\
2MASS $K_s$ & 7.288$\pm$0.024 & 15.645$\pm$0.216 & $\cdots$ &  mag & 3\\
2MASS $J$\tablenotemark{d} & $\cdots$  & 18.37$\pm$0.22 & $\cdots$ &  mag & 2\\
2MASS $H$\tablenotemark{d} & $\cdots$  & 16.66$\pm$0.22 & $\cdots$ &  mag & 2\\
MKO $J$\tablenotemark{e} & $\cdots$  &18.53 $\pm$ 0.20 &$\cdots$  &  mag & 2\\
MKO $H$\tablenotemark{e} & $\cdots$  &17.02 $\pm$ 0.20 &$\cdots$  &  mag & 2\\
MKO $K$\tablenotemark{e} & $\cdots$  &15.83 $\pm$ 0.20 &$\cdots$  &  mag & 2\\
$W1$&  7.230$\pm$0.030 & 14.461$\pm$0.014 & $\cdots$ & mag &4,5\\
$W2$&  7.285$\pm$0.019 & 13.967$\pm$0.013 & $\cdots$ & mag &4,5\\
$W3$&  7.248$\pm$0.017 & $\cdots$         & $\cdots$ & mag & 4\\
$W4$&  7.145$\pm$0.080 & $\cdots$         & $\cdots$ & mag & 4\\
ROSAT HR1 & -0.25 $\pm$ 0.23 & $\cdots$ & $\cdots$ & & 6\\
ROSAT HR2 &  0.23 $\pm$ 0.38 & $\cdots$ & $\cdots$ & & 6\\
ROSAT Count & 4.13e-02 $\pm$ 1.17e-02 & $\cdots$ & $\cdots$ & ct s$^{-1}$ & 6\\
GALEX NUV & 16.29 $\pm$ 0.022 & $\cdots$ & $\cdots$ & mag  & 7\\
GALEX FUV & 21.286 $\pm$ 0.371 & $\cdots$ & $\cdots$ & mag  &7\\
\hline
\hline
{\bf SPECTROSCOPY}\\
\hline
\hline
Spectral Type (OpT) & K0 & $\cdots$ & $\cdots$ &$\cdots$ &2\\
Spectral Type (IR) & $\cdots$ & L8$\gamma$ & $\cdots$ &$\cdots$ & 2\\
\hline
\hline
{\bf FUNDAMENTALS}\\
\hline
\hline
Age & 50 - 150 & $\le$150 & 50 - 150 & Myr & 8, 2\\
log($L_{\rm bol}$/$L_{\odot}$) & 0.35595$\pm$0.01021 & -4.33$\pm$0.09 & $\cdots$ & &9, 2\\
$T_{\rm eff}$ &  4993$\pm$124&  1303$\pm$74 & $\cdots$  & K &9, 2\\
Radius        &  0.797$\pm$0.051& 1.31 $\pm$ 0.06 & & $R_{\rm sun}$, $R_{\rm Jup}$ & 9, 2\\
Mass          &  1.0&  15$\pm$5 &  & $M_{\rm sun}$,  ~$M_{\rm Jup}$ & 10, 2\\
$\log{g}$     &  4.5539 & 4.3$\pm$0.17 & $\cdots$  & cm s$^{-2}$ & 9,2\\
\hline
\hline
{\bf KINEMATICS}\\
\hline
\hline
U     & -13.798$\pm$0.076 &$\cdots$&$\cdots$& km\,s$^{-1}$ & 2\\
V     & -28.681$\pm$0.108 &$\cdots$&$\cdots$& km\,s$^{-1}$ & 2\\
W     & -1.923$\pm$0.203 &$\cdots$&$\cdots$& km\,s$^{-1}$ & 2\\
X     & -13.923$\pm$0.010 &$\cdots$& $\cdots$ &pc & 2\\
Y     & 20.073$\pm$0.014 &$\cdots$&$\cdots$& pc & 2\\
Z     & 37.741$\pm$0.023 &$\cdots$&$\cdots$& pc & 2\\
RV & -10.15$\pm$0.24 &  $\cdots$ &  $\cdots$ & km\,s$^{-1}$ & 1\\
\hline
\hline
{\bf SYSTEM}\\
\hline
\hline
Separation&$\cdots$ & $\cdots$& 37 & $\arcsec$ & 2\\
Separation& $\cdots$ & $\cdots$& 1662 & AU & 2\\
Binding Energy & $\cdots$ & $\cdots$&1.204 & 10$^{41}$ erg  & 2\\
\enddata
\tablenotetext{a}{epoch J2016.0, ICRS}
\tablenotetext{b}{Calculated using D = 1/$\pi$, which is good approximation for parallax known to $\pi$/$\sigma_{\pi}$ = 0.75\% accuracy}
\tablenotetext{c}{Calculated using the spectrophotometric relation for young L dwarfs in \citet{Faherty16}}
\tablenotetext{d}{Synthetic 2MASS photometry from SpeX spectrum.}
\tablenotetext{e}{Synthetic MKO photometry from SpeX spectrum.}

\tablecomments{References: (1) \citet{eDR32020}, (2) This paper, (3) \citet{Cutri03}, (4) \citet{Wright10}, (5) \citet{Catwise2020}, (6) \citet{Rosat}, (7) \citet{Galex2020}, (8) \citet{Wichmann2003}, (9) \citet{TIC2018}, (10) \citet{Metchev2009} }
\end{deluxetable*}

\vspace{5mm}
\facilities{\emph{Gaia}, NASA IRTF, LICK Shane, \emph{WISE}, \emph{2MASS}}

\software{Aladin, BANYAN~$\Sigma$ (\citealt{Gagne18}), Lightkurve, SEDkit, WISEview}

\bibliography{ms.bib}
\end{document}